\documentclass[sigconf,10pt]{acmart}
\AtBeginDocument{%
  }

\setcopyright{acmlicensed}
\copyrightyear{2025}
\acmYear{2025}
\acmDOI{XXXXXXX.XXXXXXX}
\acmConference[ACM MobiCom'25]{In The 31th Annual International Conference on Mobile Computing and Networking}{Nov. 04--08,
  2025}{Hong Kong, China}
\acmISBN{978-1-4503-XXXX-X/18/06}

\usepackage{siunitx}
\usepackage{colortbl,array,xcolor}

\definecolor{myblue}{rgb}{0.5647,0.6784,0.8824}

\newcommand{\myTableCell}[1]{%
  {\renewcommand{\arraystretch}{1.0}%
  \begin{tabular}[c]{@{}l@{}}#1\end{tabular}}%
}

\begin{document}

\title[Full-body NFC]{Full-body NFC: body-scale near-field sensor networks with machine-knittable meandered e-textiles}

\author{Ryo Takahashi}
\authornote{Both authors contributed equally to this research.}
\affiliation{%
  \institution{The University of Tokyo}
  \city{Tokyo}
  \country{Japan}
  }
\email{takahashi@akg.t.u-tokyo.ac.jp}
\orcid{0000-0001-5045-341X}

\author{Changyo Han*}
\affiliation{%
  \institution{The University of Tokyo}
  \city{Tokyo}
  \country{Japan}
  }
\email{hanc@nae-lab.org}
\orcid{0000-0002-9925-3010}

\author{Wakako Yukita}
\affiliation{%
  \institution{The University of Tokyo}
  \city{Tokyo}
  \country{Japan}
  }
\email{yukita@bhe.t.u-tokyo.ac.jp}
\orcid{0009-0003-4043-2318}

\author{John S. Ho}
\affiliation{%
  \institution{National University of Singapore}
  \country{Singapore}
  }
\email{johnho@nus.edu.sg}
\orcid{0000-0003-1546-8864}

\author{Takuya Sasatani}
\affiliation{%
  \institution{The University of Tokyo}
  \city{Tokyo}
  \country{Japan}
  }
\email{sasatani@akg.t.u-tokyo.ac.jp}
\orcid{0000-0003-2268-6106}

\author{Akihito Noda}
\affiliation{%
  \institution{Kochi University of Technology}
  \city{Kochi}
  \country{Japan}
  }
\email{noda.akihito@kochi-tech.ac.jp}
\orcid{0000-0002-6393-3196}

\author{Tomoyuki Yokota}
\affiliation{%
  \institution{The University of Tokyo}
  \city{Tokyo}
  \country{Japan}
  }
\email{yokota@ntech.t.u-tokyo.ac.jp}
\orcid{0000-0003-1546-8864}

\author{Takao Someya}
\affiliation{%
  \institution{The University of Tokyo}
  \city{Tokyo}
  \country{Japan}
  }
\email{someya@ee.t.u-tokyo.ac.jp}
\orcid{0000-0003-3051-1138}

\author{Yoshihiro Kawahara}
\affiliation{%
  \institution{The University of Tokyo}
  \city{Tokyo}
  \country{Japan}
  }
\email{kawahara@akg.t.u-tokyo.ac.jp}
\orcid{0000-0002-0310-2577}

\renewcommand{\shortauthors}{Ryo Takahashi}

\begin{abstract}
Wireless body networks comprising battery-free on-body sensors and textile-based wireless readers can enable daily health monitoring and activity tracking by continuously monitoring physiological signals across the body.
However, previous textile-based wireless networks made of coils or antennas have limited the data and power transmission area because covering the whole body results in undesirable levels of electromagnetic interactions with the body, degrading the scalability, power consumption, and data rate.
Here, we report Full-body NFC, digitally-knitted electronic textiles based on a twin meander coil design that enables body-scale near-field communication (NFC) with battery-free sensor tags arbitrarily placed around the body.
Full-body NFC features i) a meander coil that enhances the magnetic field intensity on the body's surface while suppressing undesired interactions with deep tissues, in addition to ii) paired identical coil structure that enables highly-sensitive and motion-robust NFC using a differential architecture.
Additionally, industrial digital knitting machines loaded with conductive yarn allow the integration of the Full-body NFC system into daily garments supporting approximately $70-80\%$ large-scale NFC-enabled area of the body.
We demonstrate Full-body NFC could achieve mW-class energy-efficient near-field sensor networks with hundreds of kbps-class NFC battery-free sensor tags occupying less than $0.3\%$ of the coverage area under severe body movements.
\end{abstract}



\begin{teaserfigure}
  \includegraphics[width=\textwidth]{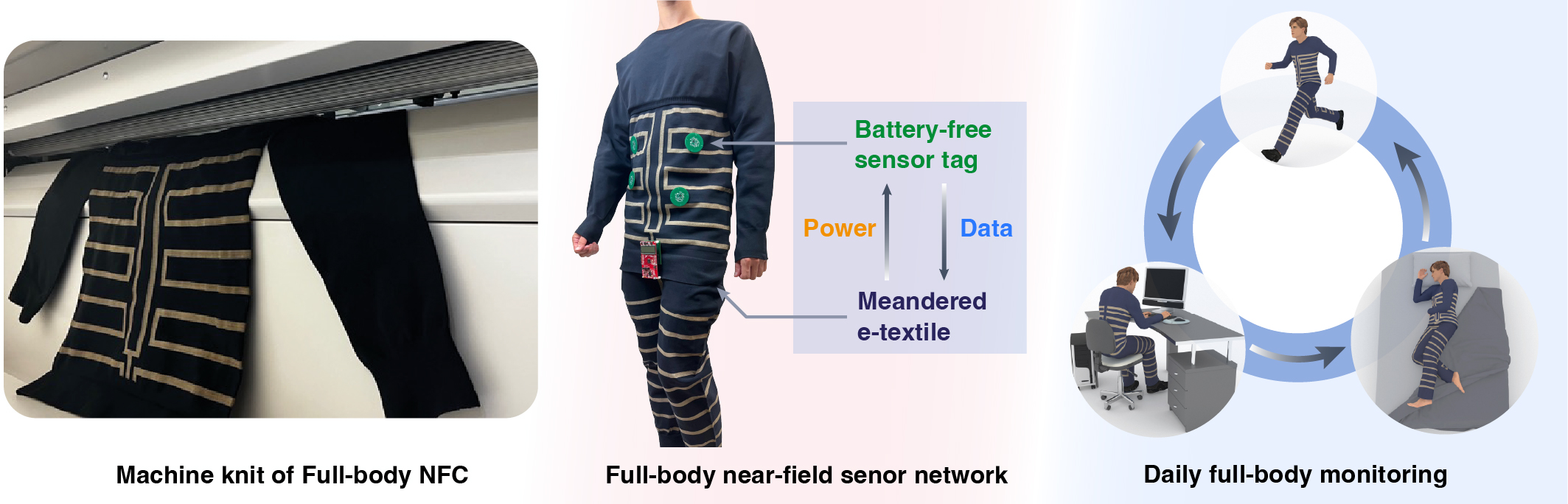}
  \caption{Overview of Full-body NFC, capable of wirelessly reading out battery-free NFC sensor tags placed throughout our NFC-enabled textiles. The textiles are automatically fabricated using industrial knitting machines. Check the demonstration video of Full-body NFC in \textcolor{blue}{\url{https://youtu.be/6x79UwY6rv8}}.}
  \label{fig:teaser}
\end{teaserfigure}

\received{20 February 2007}
\received[revised]{12 March 2009}
\received[accepted]{5 June 2009}

\maketitle

\section{INTRODUCTION}

Advances in on-body sensor networks enable unobtrusive and convenient fitness tracking and health monitoring~\cite{negra_wireless_2016}.
Because many physiological processes of interest are inherently distributed across the body, the next important step is to seamlessly interface multiple on-body sensors placed across the body.
Near-field communication~(NFC) using textile-based wireless reader coils~\cite{lin_digitally-embroidered_2022,zhu_robust_2024,lin_wireless_2020} can power up and read out battery-free sensor coil~(\textit{i.e.,} NFC sensor tag) via a near-field inductive coupling link.
The main advantage of the NFC-based near-field inductive approach~\cite{tao_magnetic_2023,whitmire_aura_2019,zhang_nfcapsule_2023} is its less electromagnetic~(EM) interaction with the dielectric human body compared to radio frequency~\cite{tian_wireless_2019,natarajan_link_2009,chen_movi-fi_2021} or body channel-based approaches~\cite{kong_power-over-skin_2024,li_body-coupled_2021,shukla_skinnypower_2019}.
However, implementing NFC-enabled electronic textiles~(e-textiles) at a body scale is still challenging.
This is mainly because scaling up the NFC reader coil near the body weakens the inductive coupling with the NFC sensor tags relative to the undesired EM interactions with the body, resulting in inefficient charging and reduced sensitivity when reading out NFC tags.
Previous NFC-enabled e-textiles have therefore been limited to narrow coverage of the body~($5-10\%$) to suppress the EM interference with the body~\cite{lin_digitally-embroidered_2022,zhu_robust_2024,lin_wireless_2020,ye_body-centric_2022}.
Consequently, such a narrow range poses availability limitations in the NFC-enabled e-textiles, requiring the redesign of the NFC reader coil for each physiological signal under interest.

\begin{table*}[t!]

{\centering
\caption{Technical comparison of wireless battery-free body sensor networks.}\label{tab:comparison}
\renewcommand{\arraystretch}{1.3}
\begin{tabular}{l|lllll}\toprule
\textbf{\myTableCell{Wireless \\body networks}} & \textbf{\renewcommand{\arraystretch}{1.}\begin{tabular}[c]{@{}l@{}}Body coverage\\percentage (\%)\end{tabular}} & \textbf{\renewcommand{\arraystretch}{1.}\begin{tabular}[c]{@{}l@{}}Wirelessly received\\power (mW)\end{tabular}} & 
\textbf{\renewcommand{\arraystretch}{1.}\begin{tabular}[c]{@{}l@{}}Sensor\\operation\end{tabular}} & \textbf{\renewcommand{\arraystretch}{1.}\begin{tabular}[c]{@{}l@{}}Datarate\\(kbps)\end{tabular}} \\\hline\hline

\rowcolor{myblue!20}UHF RFID~\cite{bouhassoune_wireless_2019,hu_biotag_2022,wu_heart_2025} & -  & $<0.1$  & {\renewcommand{\arraystretch}{1.}\begin{tabular}[c]{@{}l@{}}Within a few meter\\from external readers\end{tabular}} & $40-640^{1}$\\

BLE-enabled metamaterial~\cite{tian_wireless_2019,tian_implant--implant_2023} & 20-30  & $\mathbf{3.5}$  & \textbf{Continuous} & $100-2000^{2}$\\

\rowcolor{myblue!20}Partial-body NFC~\cite{lin_digitally-embroidered_2022,zhu_robust_2024,lin_wireless_2020,hajiaghajani_textile-integrated_2021} & 5-10  & $\mathbf{5-10}$ & \textbf{Continuous}  &$10-848$\\

Intrabody communication~\cite{kong_power-over-skin_2024,li_body-coupled_2021,shukla_skinnypower_2019} & $\mathbf{100}$ & $0.001-0.01^{3}$ & Intermittent & $-^{4}$\\

\rowcolor{myblue!20}\textbf{Full-body NFC} & $\mathbf{70-80}$ & $\mathbf{>2}$ & \textbf{Continuous} & $100-848$\\\bottomrule
\end{tabular}}\\
\begin{flushleft}
$^{1}$: ISO/IEC 18000. $^{2}$: Bluetooth 5. $^{3}$: The received power decreases inversely with increased distance from the power source. \\$^{4}$: Intrabody communication requires the equipment of battery-driven active signal transmission circuit with the sensor tag.
\end{flushleft}
\end{table*}

This paper presents \textbf{\textit{Full-body NFC}, an industrially-fabricated meandered e-textile that enables body-scale operation of arbitrarily-placed NFC sensor tags}~(see \autoref{fig:teaser}).
Full-body NFC employs a textile-based meander coil that confines the EM field near the body surface unlike conventional helical or spiral coils, thus enhancing communication sensitivity and wireless power performance.
Specifically, by implementing a ``twin'' set of identical meander coils and using them as impedance references for each other, the twin meander coil can provide a wide-band impedance balance that allows sensitive NFC despite user's dynamic motion. 
Moreover, the zigzag simple pattern of Full-body NFC is compatible with the digital knitting of conductive yarns, facilitating industrial e-textile production for full-body near-field sensor networks.
With these configurations, Full-body NFC allows energy-efficient wireless communication and charging with small NFC sensor tag occupying less than $0.3\%$ of the coverage area of approximately \SI{8000}{cm^2}.
As a proof of concept, we experimentally demonstrated that the tops- and bottoms-shaped prototype of Full-body NFC could stably readout the hundreds of kbps-class NFC sensor tag with low input power of approximately \SI{0.1}{\mW} while the NFC tag recovering enough operational power of over \SI{2}{\mW} at the input power of \SI{100}{\mW}.
Therefore, Full-body NFC capability allows for the application of the low-powered and continuous monitoring of daily biosignal such as temperature and posture through multiple NFC sensor tags arbitrarily placed across the body.
Our contributions are summarized as follows:
\begin{itemize}
    \item [1.] To the best of our knowledge, Full-body NFC is the first body-scale near-field sensor networks capable of efficiently operating arbitrarily placed NFC sensor tags around the body.
    \item [2.] Digital knitting of Full-body NFC can facilitate industrial e-textile production for full-body near-field sensor networks.
    \item [3.] We evaluated wireless power and data transmission capability of Full-body NFC under major body movements in addition to demonstration of NFC-based continuous biomonitoring. 
\end{itemize}

\section{RELATED WORK}

\subsection{Wireless Body Sensor Networks}

Body sensor networks are a network of on-body sensing devices around the human body, capable of fine-grained physiological monitoring during daily lives~\cite{gravina_wearable_2021,derogarian_miyandoab_multifunctional_2020}.
Direct wiring of these devices through the flexible PCBs~\cite{wicaksono_tailored_2020} and conductive fabrics~\cite{Noda2019Inter-ICTextile,li_plug-n-play_2025} is popular and easy to deploy.
However, the wire restricts the sensor placement on the fabric, while the vital parameters such as body temperature, heartrate, and sweat can be more accurately measured when sensors are placed directly on the skin. 
This limitation in sensor placement can lead to less precise monitoring and data collection unless the fabric is in close contact with the skin~\cite{talpur_validation_2019}.
Alternatively, wireless body sensor networks provide unconstrained connection for on-body sensors, allowing the sensors to be placed directly on the skin for more accurate physiological monitoring~\cite{niu_wireless_2019}. 
So far, a number of wireless body sensor networks have been explored utilizing various wireless technologies: ultra-high frequency radio-frequency identification (UHF RFID)~\cite{bouhassoune_wireless_2019,hu_biotag_2022,wu_heart_2025}, Bluetooth Low Energy (BLE)~\cite{tian_wireless_2019,tian_implant--implant_2023}, NFC~\cite{lin_digitally-embroidered_2022,zhu_robust_2024,lin_wireless_2020,hajiaghajani_textile-integrated_2021,noda_wearable_2019,ye_body-centric_2022}, and intrabody communication~\cite{kong_power-over-skin_2024,li_body-coupled_2021,shukla_skinnypower_2019}~(see \autoref{tab:comparison}).

Due to the small size and limited battery life of on-body sensors, there are strict power constraints on the wireless communication technologies.
UHF RFID based on RF backscatter technology allows low-powered RFID sensor tags to operate around the body without a battery by harvesting energy from an external RFID reader~\cite{bouhassoune_wireless_2019,hu_biotag_2022,wu_heart_2025}. 
Basically, the battery-free tags need to be within a few meters of the reader antenna to receive enough energy in addition to requiring angle alignment based on the specific directivity of the antenna, restricting the activation area of the UHF-RFID-based body sensor networks.
Unlike radiative transmission, BLE-enabled metamaterial antennas~\cite{tian_wireless_2019,tian_implant--implant_2023} and NFC-enabled spiral coil~\cite{lin_digitally-embroidered_2022,lin_wireless_2020,ye_body-centric_2022} are designed to confine electromagnetic waves in close proximity to the antenna and coils.
This allows for the integration of these components into clothing, enabling continuous and efficient data and power transmission for the battery-free sensor tags, while avoiding electromagnetic interaction with the body.
However, increasing the size of the antenna and coil leads to decreased efficiency and sensitivity as a result of undesirable-level electromagnetic interactions~\cite{takahashi_twin_2022}.
Intrabody communication using the conductive body as a capacitive communication medium can extend the networking area up to full body~\cite{varga_designing_2018}.
While the \si{\uW}-class power delivery across the body is possible, battery-free design of intrabody communication is challenging due to significant signal attenuation of the lossy body, requiring a battery-powered active communication module for reliably sending data throughout the entire body~\cite{kong_power-over-skin_2024, shukla_skinnypower_2019}.
Full-body NFC differs from prior works by introducing the first near-field body-scale sensor networks based on the meandered e-textiles, which enables body-scale and continuous operation of on-body battery-free NFC sensor tags with sufficient power delivery and kbps-class efficient data transmission under severe body movements.

\subsection{Backscatter Communication}

Backscatter technology, available in UHF RFID and NFC systems, is one of promising approaches for maintaining wireless on-body sensors in a low-power manner~\cite{zhao_nfc_2020,iyer_airdropping_2020,akbar_underwater_2023,xie_enabling_2024,iyer_living_2019,zhang_hitchhike_2016,zhang_enabling_2016,vasisht_-body_2018,li_internet--microchips_2020}.
The backscatter sensor tag, which transfers the data by reflecting incoming radio frequency (RF) signals, can significantly save the communication power without the need of active signal transmission. 
Additionally, either wireless power transmission~\cite{noda_wearable_2019,zhang_nfcapsule_2023,lin_digitally-embroidered_2022,zhao_nfc_2020} or energy harvesting~\cite{tao_magnetic_2023,yu_magnetoelectric_2022,jiao_zeroecg_2024}  allows the battery-free operation of the backscatter sensor tag, supporting comfortable attachment of the flexible backscatter tag to the skin~\cite{niu_wireless_2019,kim_strain-invariant_2024}.
NFC, leveraging magnetic field-based backscatter, is particularly suitable for establishing energy-efficient and stable networks around the body, owing to its minimal electromagnetic interaction with the body's dielectric properties~\cite{wang_locating_2023,ye_body-centric_2022,jiang_smart_2020,an_one_2021}. 
Full-body NFC builds on previous NFC literature by extending the near-field sensor networking area up to body scale.

\subsection{Fabrication of Electronic Textile}

The field of e-textiles has seen significant advancements in recent years, driven by the need for flexible, wearable, and integrative electronic systems.
Traditional textile manufacturing techniques such as weaving~\cite{shi_large-area_2021,poupyrev_project_2016, Noda2019Inter-ICTextile}, knitting~\cite{takahashi_meander_2022,takahashi_twin_2022,li_plug-n-play_2025,kanada_joint-repositionable_2024}, and embroidery~\cite{lin_digitally-embroidered_2022, lin_wireless_2020} have been explored to incorporate yarn-based sensors~\cite{poupyrev_project_2016}, actuators~\cite{albaugh_digital_2019}, and displays~\cite{shi_large-area_2021} into the textiles.
Machine knitting, which automatically creates a series of interlocking loops from a continuous strand of yarn, is particularly effective in creating large-scale, stretchable, breathable e-textiles~\cite{takahashi_twin_2022}. 
The inherent stretchability of knit fabric is owing to its looped structure, which allows the yarn to move and expand under tension, providing elasticity and comfort~\cite{takahashi_twin_2022,albaugh_digital_2019}.
Therefore, machine knitting method not only ensures uniformity and repeatability but also maintains the stretchablility and breathability of the fabric.
Building on the previous approaches, Full-body NFC employs machine knitting of conductive yarns to implement the body-scale, comfortable, meandered e-textiles.

\section{FULL-BODY NFC}
\label{sec:design}

\begin{figure*}[t!]
  \centering
  \includegraphics[width=\textwidth]{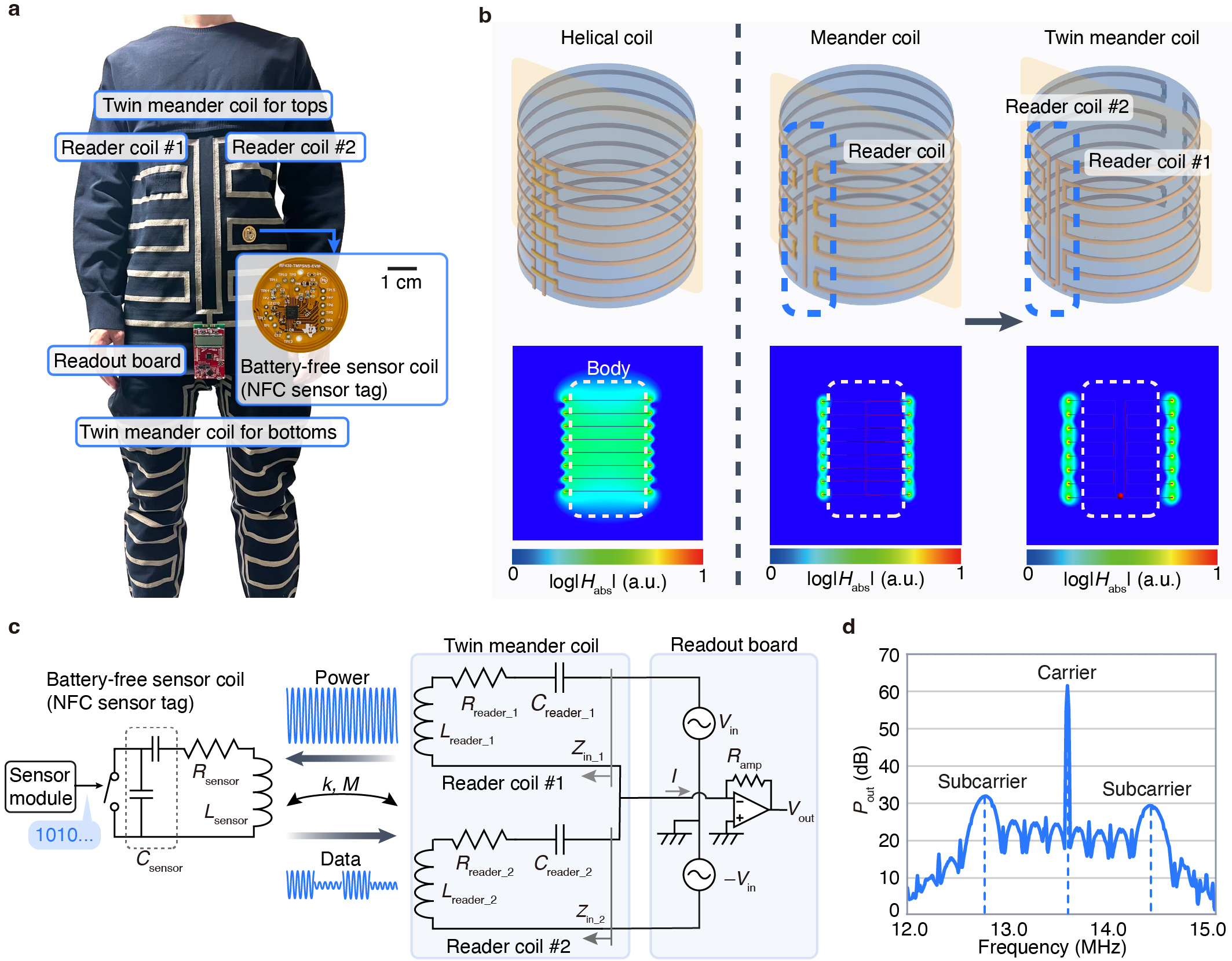}
  \caption{Design overview of Full-body NFC. 
  (a) Photograph of prototype.
  (b) Schematic and simulated inductive field of helical coil, meander coil, and twin meander coil that divides a body-scale meander coil into two parts symmetrically~(i.e., reader coil \#1 and \#2).
  (c) Circuit diagram and 
  (d) frequency spectrum of Full-body NFC.
  }
  \label{fig:design}
\end{figure*}

\subsection{System Overview}

\autoref{fig:design}a presents system overview of Full-body NFC, which consists of two reader coils forming a twin meander coil, a readout board connected to the reader coils via magnet connectors, and a battery-free sensor coil including a sensor module.
The readout board is mainly composed of a balanced bridge circuit~\cite{takahashi_picoring_2024,takahashi_telemetring_2020}.
When the sensor coil is placed near the reader coils, an inductive field~(\SI{13.56}{\MHz} industrial, scientific, and medical radio band) generated by either of the reader coils couples with the sensor coil to construct an inductive coupling.
Through the inductive coupling, the reader coil sends \si{\mW}-class operational power to the sensor coil. 
Then, the sensor coil powers up the sensor module and encodes the digitized sensor information by changing its impedance with ON/OFF switching.
The switching by the sensor coil can induce a change in the input impedance of the reader coil, $Z_{\rm in}$ as follows:
\begin{align}
Z_{\rm in} &= Z_{\rm reader} + \Delta Z_{\rm reader}\\
\Delta Z_{\rm reader} &= \cfrac{(2\pi fM)^2}{Z_{\rm sensor}}
\end{align}
where $f$ the readout frequency, $M=k\sqrt{L_{\rm reader}L_{\rm sensor}}$ the mutual inductance between the sensor and reader coil, $k$ inductive coupling coefficient between the reader and sensor coils, and $Z_{\rm sensor}$ the impedance of the sensor coil.
The change in $Z_{\rm in}$ can be decoded using the readout board.
Because the sensor coil can receive \si{\mW}-class power from the reader coil, the battery-free operation of the sensor coil is possible~\cite{zhao_nfc-wisp_2015,hajiaghajani_textile-integrated_2021,lin_digitally-embroidered_2022}.

\subsection{Sensitive NFC by Twin Meander Coil}

\begin{figure}[t!]
  \centering
  \includegraphics[width=1\columnwidth]{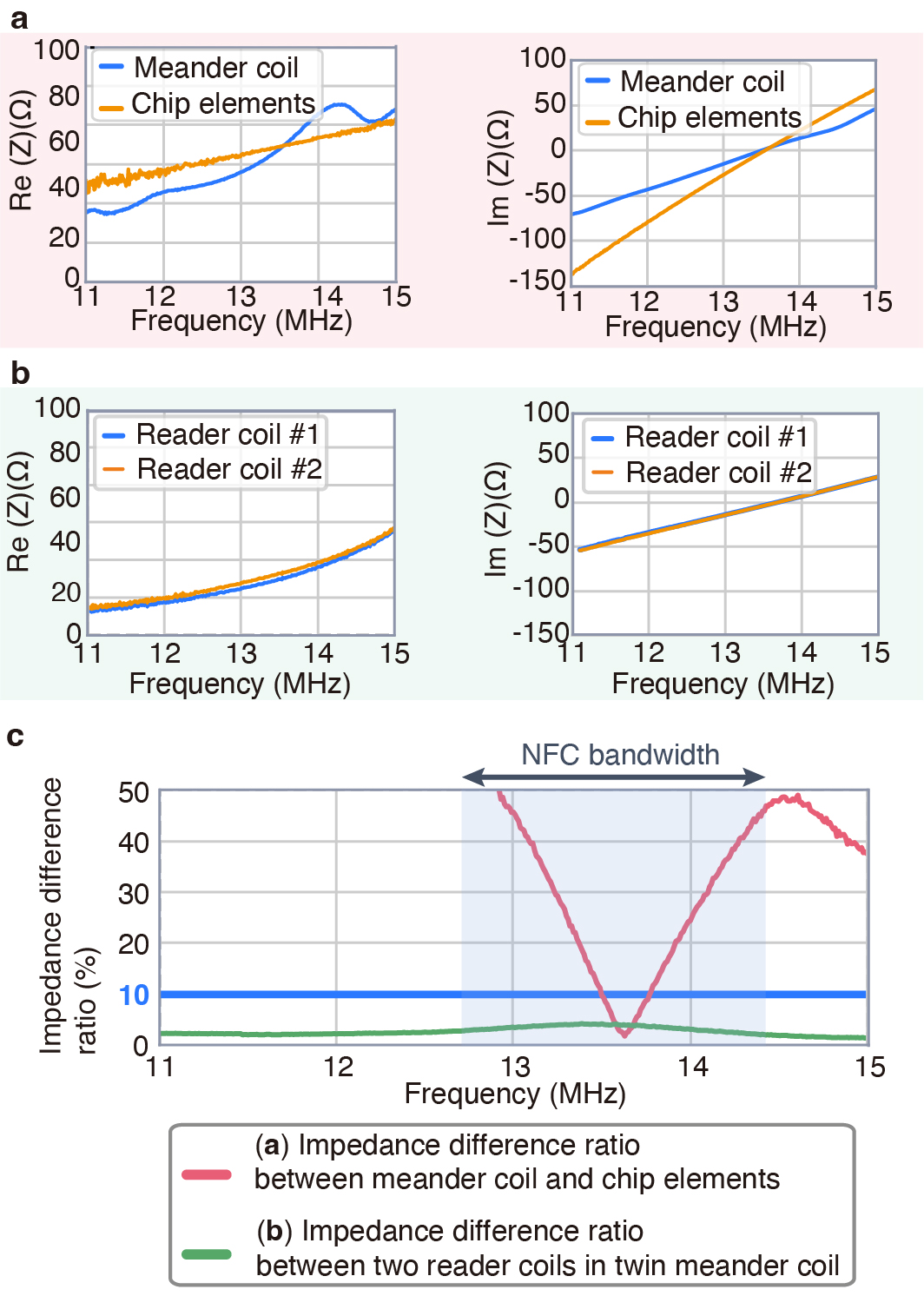}
  \caption{
  Measured impedance characteristics of two types of body-scale reader coil: (a)~body-scale meander coil and chip elements impedance-matched with the reader coil at 13.56 MHz and (b) twin meander coil consisting of the two identical meander coil structure.
  (c)~Impedance difference ratio in (a) and (b). The impedance-balanced frequency band is here below 10\% of the impedance difference ratio.}
  \label{fig:impedance}
\end{figure}

A major challenge of the NFC-enabled e-textile system is that the reader coil needs to be as small as the sensor coil to construct a strong inductive coupling between the sensor and reader coil~($k>0.1$)~\cite{lin_digitally-embroidered_2022,lin_wireless_2020,ye_body-centric_2022}.
When a reader coil such as a helical or spiral coil extends up to a body scale, the inductive coupling becomes significantly weak~($k<0.005$) and unstable because of the large size ratio of the body-scale reader coil to the small sensor coil, in addition to the inbody strong electromagnetic interaction caused by the body-scale reader coil.
Therefore, $\Delta Z_{\rm reader}$ becomes approximately below \SI{10}{\mohm}~(the change ratio of $Z_{\rm in}$ below $0.1\%$); such a small change in $Z_{\rm in}$ cannot be read out.

To enable sensitive readout of small sensor coils via a body-scale reader coil, we employ a twin meander coil approach~\cite{takahashi_twin_2022}, which divides a body-scale meander coil into two identical parts and connects them with the balanced bridge circuit.
Unlike standard helical or spiral coils that generate a strong electromagnetic field permeating into the body when scaling up, a meander coil can confine the strong electromagnetic field around the clothing's surface by changing its winding direction for each turn~\cite{takahashi_twin_2022,takahashi_meander_2022}.
\autoref{fig:design}b shows simulation results for the inductive field in a body-scale helical coil, a meander coil, and a twin meander coil as analyzed using an electromagnetic solver called Altair Feko.
The result shows the meander coil enables the confinement of the strong inductive field near the skin's surface, unlike the helical coil.
We also measured that $k$ and $\Delta Z_{\rm reader}$ of the body-scale meander coil become approximately $0.04$ and \SI{1}{\ohm}~(the change ratio of $Z_{\rm in}$ about $5\%$).

\begin{figure*}[t!]
  \centering
  \includegraphics[width=\textwidth]{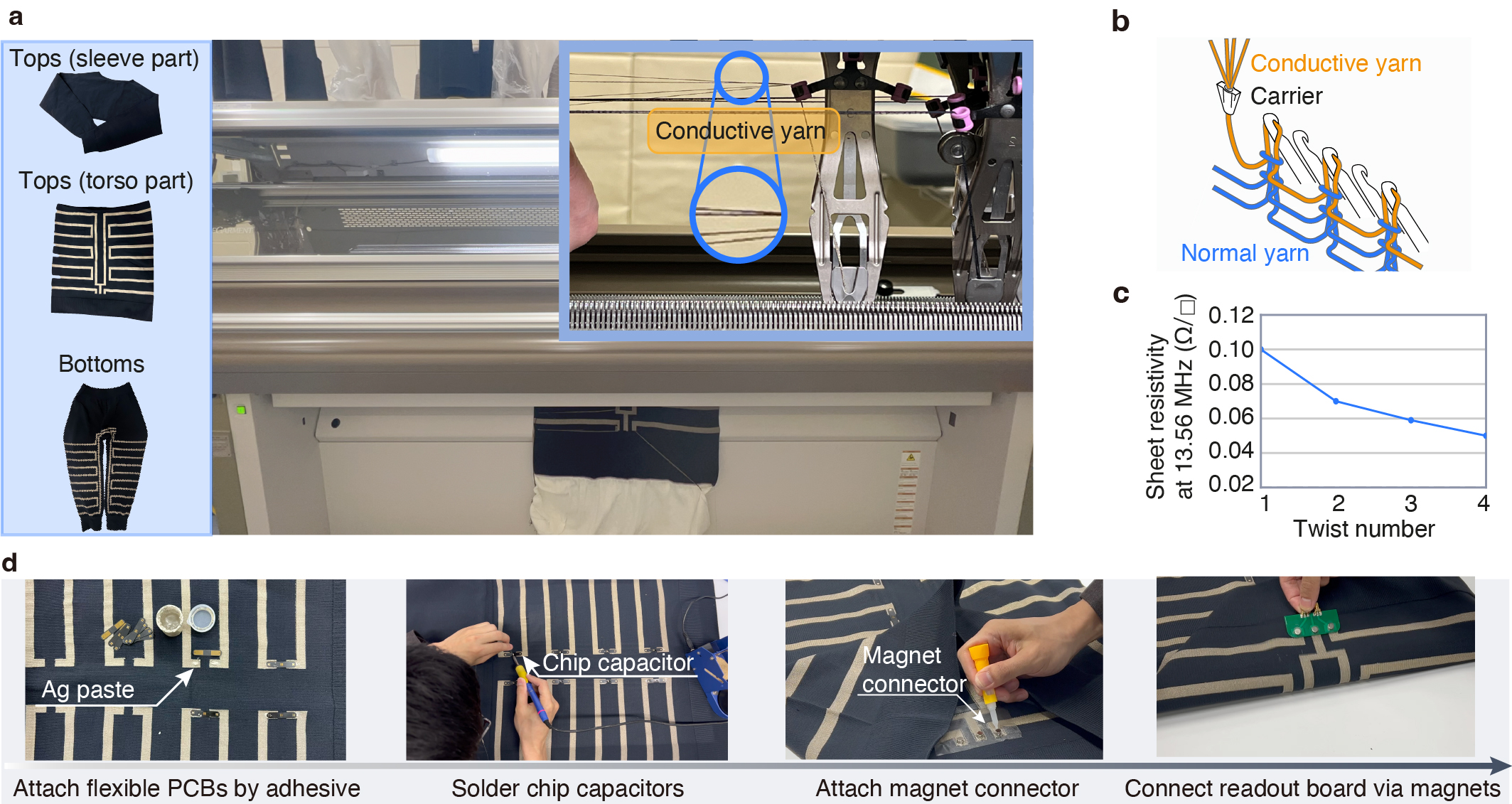}
  \caption{
  Fabrication process of Full-body NFC. 
  (a) Photograph and (b) schematic of knitting process of the twin meander coil. The knitting machine joins multiple yarns via a carrier to decrease the wire resistive characteristics.
  (c) Sheet resistivity of knitted conductive wire for a joining number of conductive yarns.
  (d) Photograph of the fabrication post-process of the twin meander coil.}
  \label{fig:fab}
\end{figure*}

Moreover, a balanced bridge circuit is used to detect the small variations of $Z_{\rm in}$ caused by the sensor coil, as shown in \autoref{fig:design}c.
The balanced bridge circuit, which comprises two input sources, one reader element~($Z_{\rm reader}$), and one reference element~($Z_{\rm ref}$), can detect the small impedance variations of the reader element by equalizing the reference element with the reader element~($Z_{\rm reader}=Z_{\rm ref}$), i.e., impedance balance~\cite{takahashi_picoring_2024,takahashi_telemetring_2020}.
However, because the impedance characteristics of the reader coil knitted by conductive threads are complicated, it cannot achieve impedance balance with the reference element, which consists of chip inductors, capacitors, and resistors.
This is because knitted textiles have many stray capacitors between the loop of the conductive yarns.
Therefore, twin meander coil~\cite{takahashi_twin_2022} has reported an approach that achieves almost perfect impedance balance of the textile coil by connecting the two identical textile coils with the balanced bridge circuit.
With this configuration, each textile coil can work as \textit{reader} while utilizing each other as a reference for impedance balance.
Note that prior twin meander coil systems require a wide impedance-balanced frequency band to concurrently read $LC$ sensors assigned at different resonant frequencies via frequency-division approach.
Such a wide band requirement causes the readout error against user movements because the deformation of the coil by the user movement narrows the impedance-balanced frequency band. 
In contrast, Full-body NFC solves this challenge by tuning the sensor coil at the same resonant frequency and communicating with the multiple tags in time-division multiplexing manner~(\textit{i.e.,} NFC).
Such a time-division approach enables the stable readout of multiple sensor coils even when the impedance-balanced frequency band becomes narrow due to the user's motions.

To validate the impedance balancing with the twin meander coil, \autoref{fig:impedance}a shows the impedance characteristics of the body-scale knitted meander coil within the frequency range of \SI{11}{\MHz} to \SI{15}{\MHz}, and the impedance characteristics of the chip elements, both of which are impedance-balanced at \SI{13.56}{\MHz}.
Moreover, \autoref{fig:impedance}b shows the impedance characteristics of the twin meander coil consisting of two meander coils~(see \autoref{fig:design}a).
The impedance-balanced frequency band, where the impedance difference ratio is below $10\%$, is \SI{0.2}{\MHz} between the body-scale meander coil and chip elements~(see \autoref{fig:impedance}c)
Such a narrow frequency band cannot cover the NFC frequency band~(\SI{13.56}{\MHz}$\pm$\SI{848}{\kHz}, see \autoref{fig:design}d). 
By contrast, the twin meander coil can accommodate the bandwidth required for the NFC protocols.

Lastly, let us explain the detail of the balanced bridge circuit with the twin meander coil.
The output of the balanced bridge circuit, $V_{\rm out}$, connected with the twin meander coil can be calculated as follows when $Z_{\rm in\_1} = Z_{\rm in\_2}$:
\begin{align}
   &V_{\rm{out}} = -R_{\rm{amp}}I =-R_{\rm{amp}}\left(\cfrac{V_{\rm{in}}}{Z_{\rm{in\_1}}}-\cfrac{V_{\rm{in}}}{Z_{\rm{in\_2}}}\right) \notag \\ 
       &\approx
    \begin{cases}
      0\mbox{:~no sensor \#i} \\
      \pm R_{\rm{amp}}\cfrac{\Delta Z_{\rm{in\_1}}}{Z_{\rm{in\_1}}^2}V_{\rm{in}}~\mbox{:~sensor coil \#i on reader coil \#1 or \#2} \\
   \end{cases} \label{eq:v_out_matched}
\end{align}
where $V_{\rm out}$ and $V_{\rm in}$ the output and input signals of the balanced bridge circuit, respectively, $R_{\rm amp}$ the amplifier factor of the inverting amplifier, $I$ the input current flowing to the inverting amplifier, and $Z_{\rm in\_1}$ and $Z_{\rm in\_2}$ are the input impedance of reader coil \#1 and \#2, respectively.
\autoref{eq:v_out_matched} shows that the balanced bridge circuit can detect the small variations $\Delta Z_{\rm in\_i}$ 
Owing to the relatively wide impedance-balanced bandwidth~(see \autoref{fig:impedance}c) and the sensitivity of the bridge circuit, Full-body NFC can reliably read out $\Delta Z_{\rm in}$ via $V_{\rm out}$ against user motions.

\section{IMPLEMENTATION}

\begin{figure*}[t!]
  \centering
  \includegraphics[width=\textwidth]{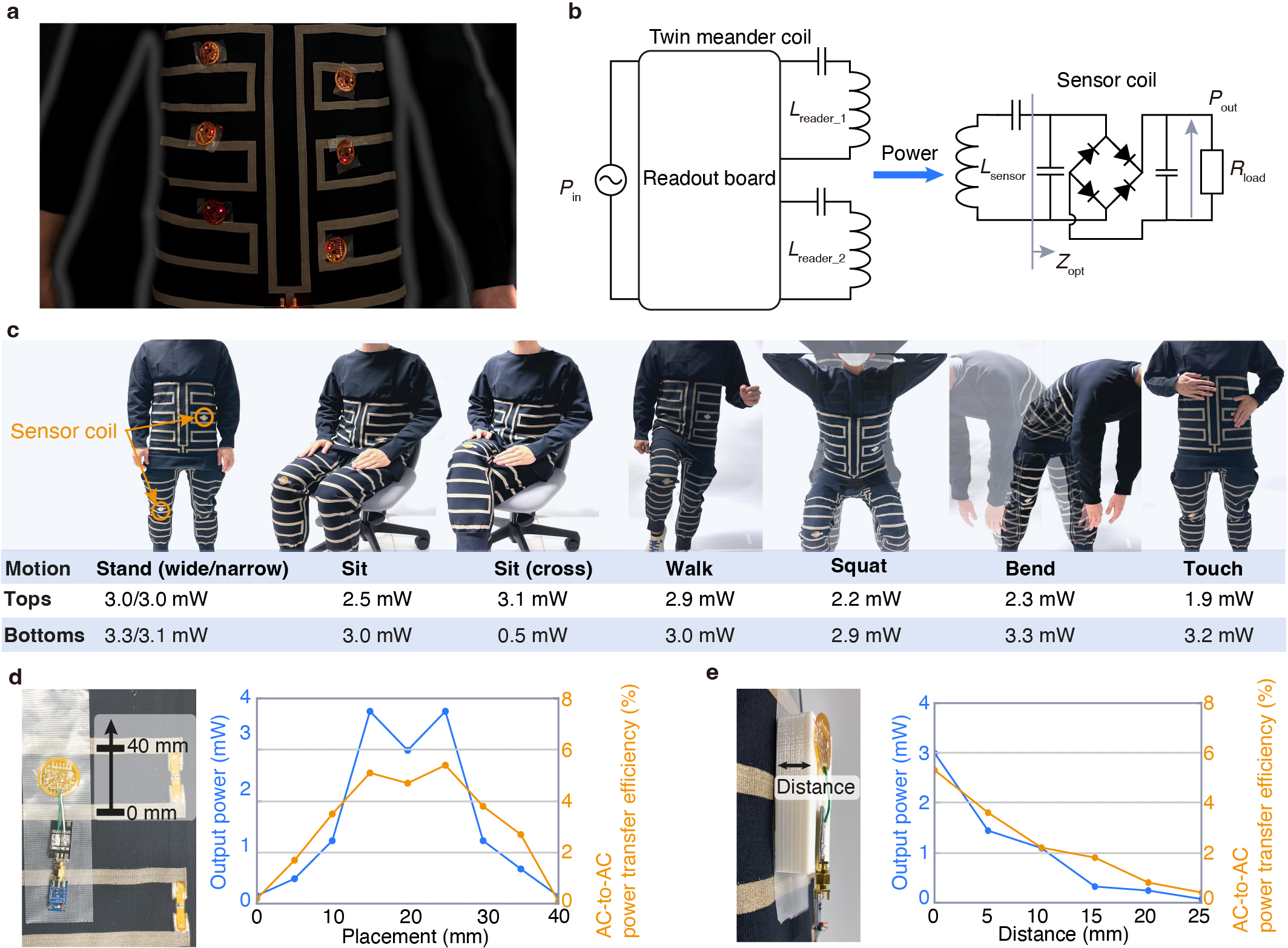}
  \caption{Experimental verification of body-scale wireless charging capability.
  (a) Photograph of wireless charging demonstration of lighting up six NFC sensor coils equipped withred LED. 
  (b) Schematic of power supply scheme using two reader coils. 
  (c) Result of output power for eight different user motions..
  (d)(e) Result of output power and AC-to-AC power transfer efficiency for (d) sensor's placement and (e) distance of the sensor coil from the reader coil.
  }
  \label{fig:power}
\end{figure*}

The fabrication process of the twin meander coil is shown in \autoref{fig:fab}.
Because the meander pattern is simple zigzag pattern, the twin meander coil can be created by digital knitting~(MACH2XS 15S, Shima Seiki).
First, we knitted two types of garments~(tops and bottoms) using a digital knitting machine of conductive yarn~(AGfit, Mitsufuji)~(see \autoref{fig:fab}a).
The tops and bottoms include a pair of the two reader coils $\#$1,2~(wire width:~\SI{1}{\cm}, wire space:~\SI{4}{\cm}), respectively.
Because the typical conductive thread is lossy, we jointed the multiple threads via a thread carrier to decrease the resistive loss of the total conductive threads~(see \autoref{fig:fab}b).
Due to the machine limitation, the maximum joint number of the yarn is $4$; the resistivity of the conductive threads joining four threads is reduced by approximately half~(see \autoref{fig:fab}c).
Subsequently, we heat-pressed the flexible PCBs on the clothing with silver paste to solder the multiple chip capacitors, which are used to tune the resonant frequency of a long coil to \SI{13.56}{\MHz}.

To make a relatively long coil considering the short wavelength at the \SI{13.56}{\MHz} band, we used a technique called distributed capacitance arrangement~\cite{cook_large-inductance_1982,takahashi_picoring_2024,takahashi_telemetring_2020}, which constructs a longer coil by splitting the long coil into the multiple short coils connected in series with multiple capacitors.
The distributed capacitor also weakens the capacitive interference with the dielectric body, thereby resulting in a stable coupling between the coils near the dielectric body.
We inserted four \SI{250}{\pF} capacitors into each reader coil of the tops, and five \SI{230}{\pF} capacitors into each reader coil of the bottoms.
In total, the measured $Q$-factor, resistance~($R$), and inductance~($L$) of the reader coil of the tops/bottoms were \num{10.4}/\num{11.1}, \SI{18}{\ohm}/\SI{23}{\ohm}, and \SI{2.2}{\uH}/\SI{3.0}{\uH} at \SI{13.56}{\MHz}, respectively.
Note that the two reader coils of the tops are placed \SI{5}{\cm} away from each other to mitigate the unnecessary inductive coupling between the two reader coils.

\section{EVALUATION}
\label{sec:evaluation}

\subsection{Wireless Charging Capability}
\label{sec:wireless_charging_capability}

First, we evaluated the wireless charging capabilities of Full-body NFC for sensor number, different body motions, distances, and sensor alignments.
\autoref{fig:power}a shows the demonstration of wireless charging for six NFC sensor tags~(diameter:~\SI{3}{\cm}, turn number:~$6$, $Q$-factor:~$34$) placed across the twin meander coil. The input power is set up to be \SI{100}{\mW}.
The twin meander coil could delivery enough power to the NFC sensor tags, resulting in light-up of a light-emitting diode~(LED) connected to the tag.
Note that \autoref{fig:power}b shows the schematic of wireless charging using the twin meander coil.
We confirmed that the maximum number of NFC sensor tags that can be powered simultaneously is $10$~(see \textcolor{blue}{\url{https://youtu.be/6x79UwY6rv8?t=107}}).

Then, we measured the average output power~($V_{\rm out}^2/R_{\rm load}$) and AC-to-AC power transfer efficiency for the three users~(1 woman, 2 men) when the optimal load~($Z_{\rm opt}$) is connected to the sensor coil~(see \autoref{fig:power}c-e).
The optimal load indicates that the sensor coil can achieve the maximum AC-to-AC power transfer efficiency~\cite{zargham_maximum_2012} when the user is standing.
\autoref{fig:power}c shows that the output power at an input power~($P_{\rm in}$) of \SI{200}{\mW} is almost the same against the various dynamic motions, indicating that power delivery by the twin meander coil is robust against coil deformation caused by body movements.
This is because the twin meander coil with a low $Q$-factor of approximately $10$ can mitigate the efficiency decrease under coil's deformation, although the original efficiency is relatively low~(5\%).
Fortunately, owing to the low power consumption of the NFC sensor coils~($<$\si{\mW}), wireless charging at such a low efficiency can stably drive the NFC sensor coils.
Note the output power of the sensor coil placed on the bottoms becomes almost $0$ during sitting with legs crossed because of the electrical contact between the two reader coils; this could be solved by covering the wire with insulators.

Lastly, we measured how the sensor misalignment and the distance between the sensor and the twin meander coil can affect the output power and AC-to-AC power transfer efficiency (see \autoref{fig:power}d-e).
Because the meander coil generates a short-range strong inductive field between the wires, the reader coil can achieve an output power of \SI{\ge1}{\mW}. 1)~within a \SI{\pm 1}{\cm} misalignment from the center between wires of the twin meander coil, and 2)~within a \SI{1}{\cm} distance from the center.

\subsection{Wireless Communication Capability}
\label{sec:wireless_communication_capability}

\begin{figure*}[t!]
  \centering
  \includegraphics[width=\textwidth]{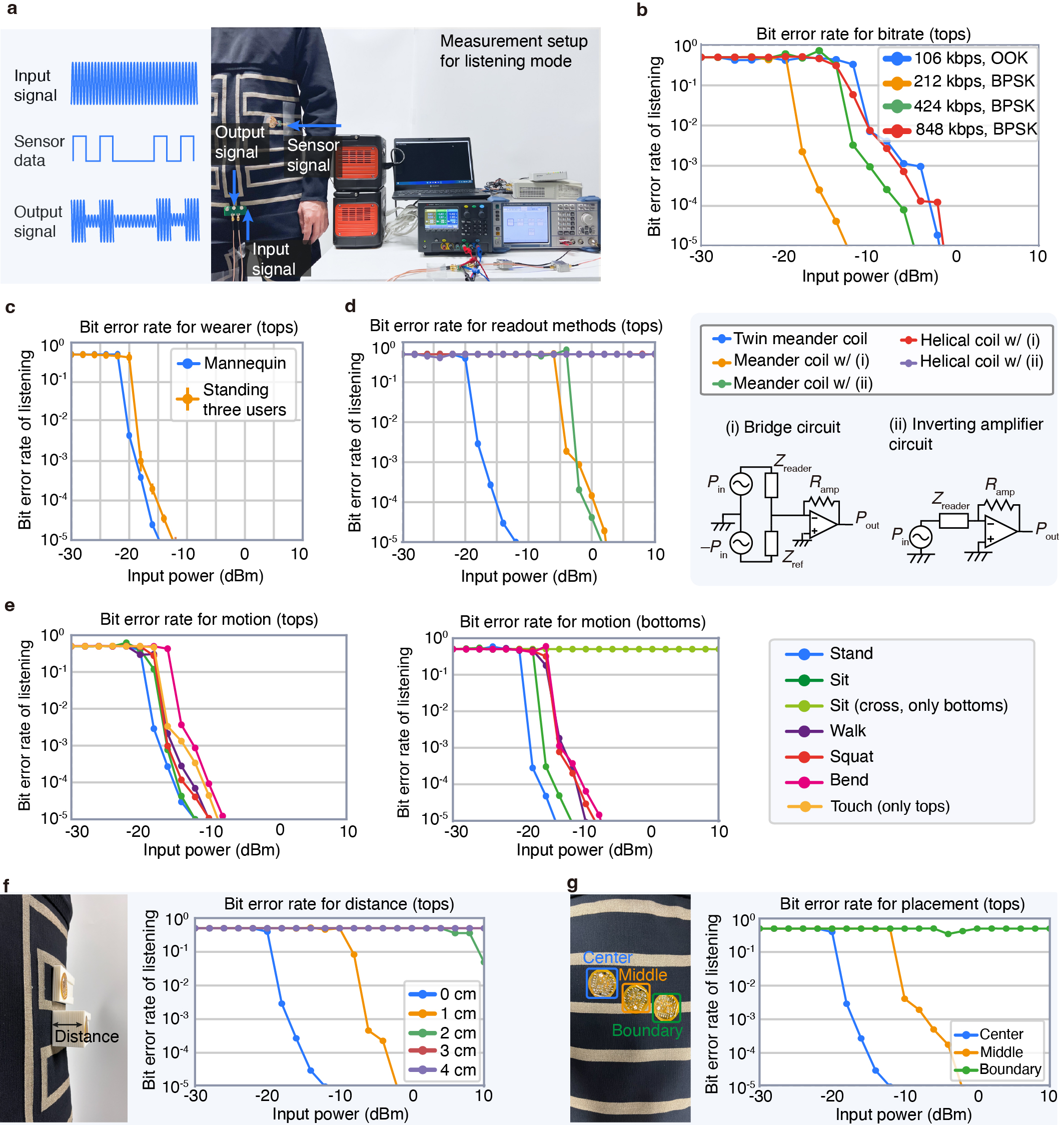}
  \caption{Experimental verification of body-scale wireless communication capability.
  (a) Configuration of bit-error-rate~(BER) measurement of the \textit{listening} mode.
  (b) Results of BER for different bit rate with different keying.
  (c) Results of BER for mannequin or human.
  (d) Results of BER of five types of readout methods.
  (e) Results of BER of the tops and bottoms for various motions. 
  (f)(g) Results of BER according to (f) distance of the sensor coil from the reader coil and (g) sensor's placement.
  }
  \label{fig:comm}
\end{figure*}

Next, we evaluated the wireless communication performance based on the NFC specifications.
NFC is designed to exchange data following the two modes below:
First, in the \textit{polling} mode, the reader coil powers up the NFC sensor coil and emits a polling signal at a \SI{13.56}{\MHz} carrier signal.
Then, in the listening mode, the reader coil receives the sensor data signal modulated by the sensor coil at the subcarriers \SI{848}{\kHz} away from the carrier frequency.
Because the wireless communication performance of the listening mode is lower than that of the polling mode because the modulated subcarriers have much lower power than the carrier, our examinations focus on the bit error rate~(BER) in the listening mode.
The BERs were measured by the ratio of the number of bits successfully received by the reader coil to the total number of transmitted bits from the sensor coil, varying the input power $P_{\rm in}$ from \SI{-30}{dBm} to \SI{10}{dBm} in steps of \SI{2}{dB}.

\begin{figure*}[t!]
  \centering
  \includegraphics[width=\textwidth]{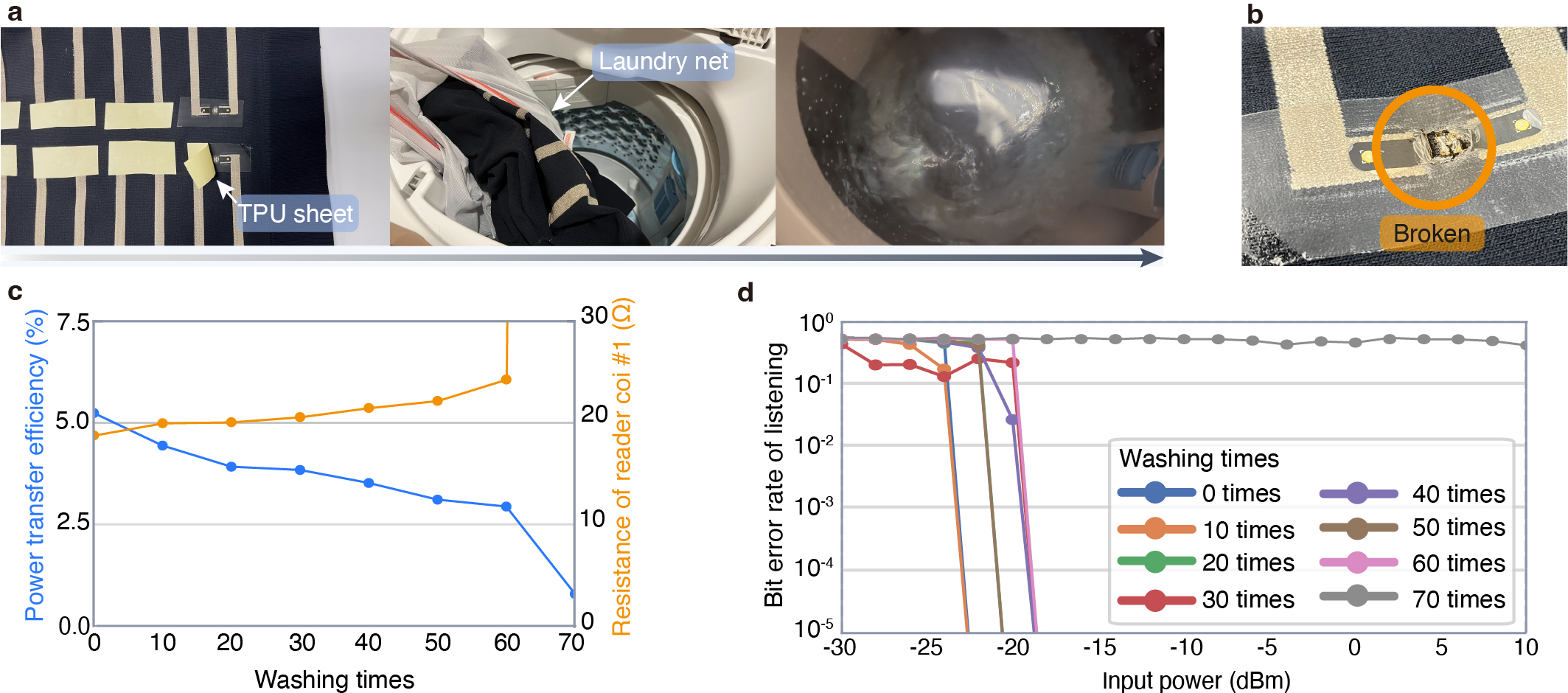}
  \caption{
  Overview of durability and wireless charging/readout capability for machine washing. 
  (a) Experimental setup of machine washing of the clothing. 
  (b) Photograph of broken flexible PCBs by machine washing of over 60 times.
  (c) Results of AC-to-AC power transfer efficiency and resistance of the reader coil \#1 for every 10 times of machine washing.
  (d) Result of BER in the \textit{listening} mode for number of machine washing.
  }
  \label{fig:wash}
  \Description{}
\end{figure*}

The entire experimental setup is depicted in \autoref{fig:comm}a.
To measure BER, the setup prepares an input signal to the reader coil, i.e., carrier signal, the sensor signal, and the output signal from the reader coil.
The input signals were generated with a vector signal generator~(SMBV100A, Rohde\&Schwarz).
The random sensor signals~(pseudorandom binary sequence 15) were generated with a custom Python script to follow the NFC-A specification.
To modulate the sensor signals into the input signal, we connect the sensor coil to the load-modulation switch comprised of two transistors~(BC847C-7-F, Diodes Incorporated).
The voltage to turn on the switch is \SI{1}{\V}.
The output signal was received with a software-defined radio platform~(USRP N210, Ettus Research).
After the reception, the output signals were offline-processed using digital signal processing in the PC connected to the USRP.
First, the signal was lowpass-filtered using an IIR Butterworth filter to eliminate unwanted signals roughly.
We then applied a decision-directed least-mean-squares equalizer to extract the symbols.
Finally, after decoding the symbols, we calculated the BER from the sequence.
To prevent both the reader and sensor coils from sharing a common ground, we used two portable power supplies and two RF transformers~(TR010-0S, R\&K Company Limited) to separate the ground.

We evaluate BER curve for different bitrates, wearers, body movements, distance and misalignment between the sensor and reader coils.
First, \autoref{fig:comm}b shows the BER curves for the four bitrates.
The NFC standard supports different bit rates with different modulation schemes.
Here, we refer to a specification compatible with ISO/IEC 14443 provided by~\cite{noauthor_high-performance_2024}, which defines four types of bitrates with two different modulation schemes: \SI{106}{kbps} with OOK, \SI{212}{kbps}, \SI{424}{kbps}, and \SI{848}{kbps} with BPSK.
Because BPSK provides better BER performance compared to OOK, the listening mode with a bit rate of \SI{212}{kbps} yields the lowest BER under the same $P_{\rm in}$.
Therefore, we used the bitrate of \SI{212}{kbps} in the subsequent experiments.
Next, \autoref{fig:comm}c illustrates the BER curve for two types of wearers: mannequin or three people~(one woman, two men). 
The result indicates the BER curve is almost the same independent of the wearers because the meander coil can avoid the electromagnetic interaction with the dielectric body.
Furthermore, comparisons were made between a balanced bridge circuit and an inverting amplifier circuit, and among a twin meander coil, a meander coil, and a helical coil.
The result indicates that the combination of the twin meander coil and the bridge circuit increases the sensitivity by approximately \SI{13}{\dB} at a BER of \num{e-3}~(see \autoref{fig:comm}d).
In addition, the twin meander coil shows a stable wireless readout performance under diverse motions, respectively, except sitting with legs crossed which results in the readout failure because of the contact between the two reader coils~(see \autoref{fig:comm}e).
We note that the misalignment and distance of the sensor from the reader also affect BER~(see \autoref{fig:comm}f-g).
Over \SI{\pm 2}{\cm} misalignment of the sensor from the center between wires of the twin meander coil or a \SI{\ge 3}{\cm} distance from the twin meander coil results in BER of \num{0.5} at even at an input power exceeding \SI{10}{dBm}.
Based on these results described in \autoref{fig:comm}b-e, the acceptable BER performance at $10^{-3}$ was estimated to be over \SI{-10}{dBm}~(\SI{0.1}{\mW}) of the input power.

\subsection{Washing Durability}

Lastly, we examine the mechanical durability of Full-body NFC against daily washing.
The clothing was inserted in a commercially available laundry net to mitigate the mechanical stress of machine washing, and was washed and dried repeatably~(see \autoref{fig:wash}a).
Note that the TPU protection sheet was covered on every flexible PCBs. 
The result shows that one of the flexible PCBs, which is used to connect chip capacitors to the meander coil, was broken by mechanical stress after $70$ cycles of machine washing, resulting in charging and communication dysfunction of Full-body NFC~(see \autoref{fig:wash}c-d).
While Full-body NFC can potentially endure for daily wear and wash to some extent (i.e., approximately a few years), the use of mechanically-durable rigid-to-stretchable connection using liquid metal~\cite{sato_method_2021,sato2025mems} could make the capacitance area more durable against numerous washing.

\section{APPLICATION EXAMPLES}
\label{sec:application}

\begin{figure*}[t!]
  \centering
  \includegraphics[width=\textwidth]{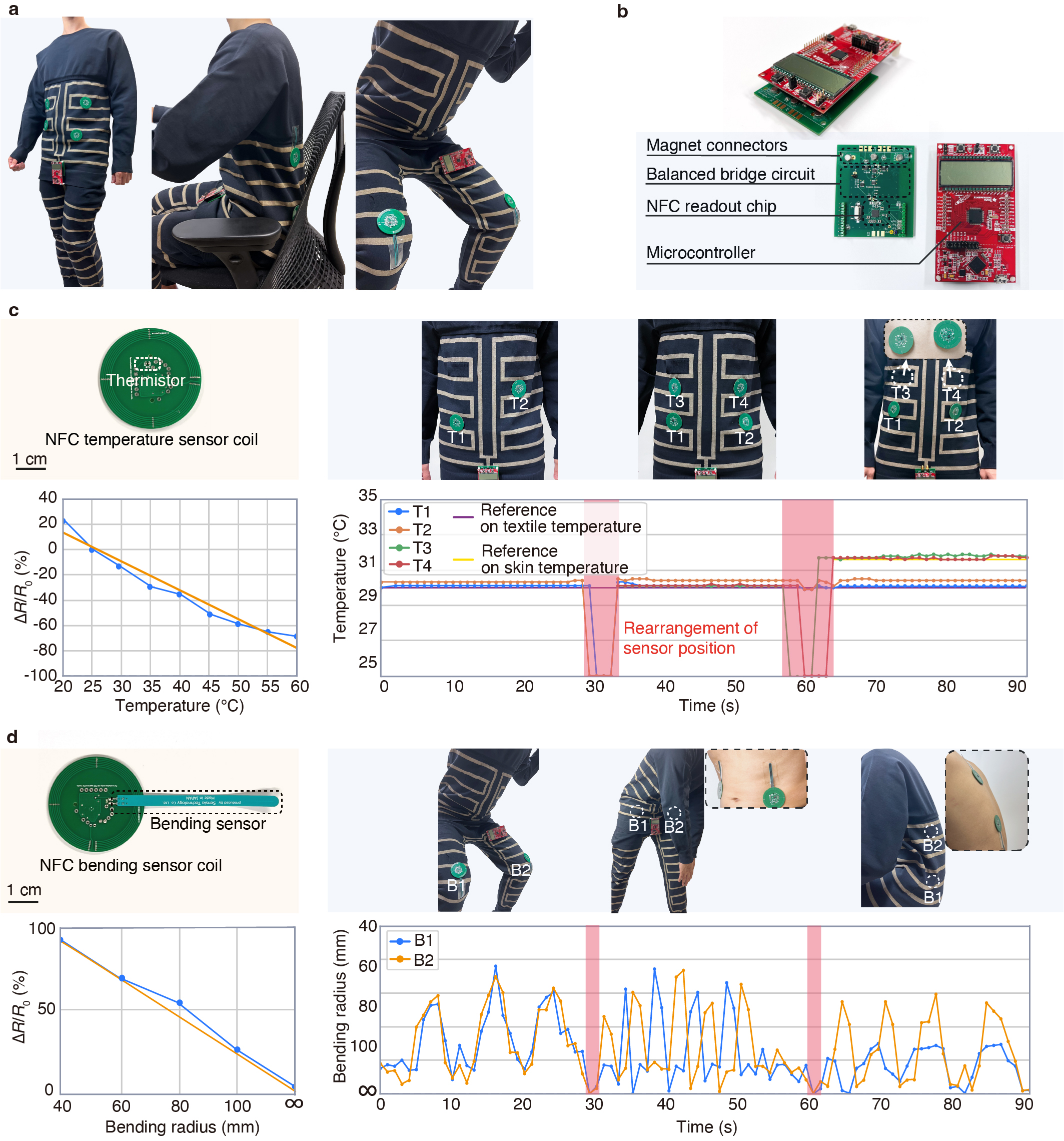}
  \caption{
  Application examples of Full-body NFC.
  Photograph of (a) the Full-body NFC prototype and (b) NFC readout board consisting of magnet connectors, balanced bridge circuit, NFC readout chip, and microcontroller.
  (c) Temperature mapping with four NFC temperature sensor coils placed on the tops during the user's walking.
  (d) Motion sensing with two NFC bending sensor coils attached to either tops or bottoms during the user's bending of the knee, abdomen, and back.
  }
  \label{fig:app}
  \Description{}
\end{figure*}

To demonstrate the wireless charging and readout utility of Full-body NFC, we demonstrated NFC-based full-body monitoring with NFC sensor coils.
\autoref{fig:app}a-b show the prototype of our NFC-based meandered electric textiles using a custom NFC-enabled readout circuit.
The readout board consists of an NFC readout chip (TRF7970A, Texas Instruments) controlled by a microcontroller~(MSP-EXP430FR4133, Texas Instruments), the balanced bridge circuit, and three neodymium magnet connectors with a diameter of \SI{5}{\mm} and a height of \SI{3}{\mm}.
The magnet connectors are used to strongly connect the readout board to the magnet connectors attached to the back side of the twin meander coil.
The NFC sensor coils are equipped with an off-the-shelf NFC transponder integrated with temperature or bending sensors and a rigid resonant coil tuned at \SI{13.56}{\MHz}~($k:~0.05$, $L$:~\SI{3.0}{\uH}) with a diameter of \SI{4}{\cm} and a turn number of $6$.
To begin the NFC communication with the off-the-shelf NFC readout chip, we must increase $\Delta V_{\rm out}$ over \SI{1}{\V} by increasing both the diameter of the sensor coil used in \S~\ref{sec:wireless_communication_capability}.

\autoref{fig:app}c-d show two application examples of NFC-based full-body monitoring integrated with temperature or bending sensors.
Based on time-domain multiple access, the reader at an input power of \SI{200}{\mW} records resistance data from up to four or five NFC sensor coils at a sampling rate of $1-2$~\si{\Hz} per sensor.
Here, two or four NFC sensor coils were located on various placements of the tops or bottoms to monitor the garment or skin temperature during the user's waking. 
The representative data of each sensor during short-term exercise is shown on the right side of \autoref{fig:app}c-d.
The NFC sensor data was estimated based on the least-squared line relationship between the actual resistance change ratio of the NFC sensor coil for applied temperature or bending radius~(see the left side of \autoref{fig:app}c-d).
As for the temperature monitoring, the reference temperature was also measured by an infrared camera or an RGB camera. 
Due to the packet collision, our current NFC demonstration cannot support the readout of the NFC sensor tags over five.

\autoref{fig:app}c illustrates that the recorded temperature data of the NFC sensor coils matches well with the thermal camera-based reference textile temperature~($\sim\SI{30}{\degreeCelsius}$) with small errors~($\SI{<0.5}{\degreeCelsius}$) during $1$ minute.
The sensor coils can also detect the slight increase of approximately \SI{2}{\degreeCelsius} when the location of the two sensor coils~(T3, T4) changed from the textile to the skin.
Then, \autoref{fig:app}d shows that the output of the two bending NFC sensor coils corresponds to the change in the bending radius of the user's body during three different user's bending motions including bending and squatting.
Full-body NFC is able to stably capture the sensor's data without almost packet loss against the user's motion~(see \textcolor{blue}{\url{https://youtu.be/6x79UwY6rv8}}).
Such motion-robust temperature mapping and motion recording are useful in daily clinical and athletic applications as the evaluation methods of textile breathability, mental stress, or athletic performance for various daily situations~\cite{libanori_smart_2022}.
Unlike the previous narrow-area NFC-enabled clothing~\cite{lin_digitally-embroidered_2022,lin_wireless_2020}, Full-body NFC can support multi-purpose full-body sensing during daily lives without major modification of the reader coil pattern and arrangement.
Therefore, Full-body NFC has the potential to be a versatile platform for full-body monitoring, by simply relocating appropriate NFC sensor coils according to target application scenarios.

\section{DISCUSSION AND CONCLUSION}

We have reported Full-body NFC, a digitally-knitted twin meander coil that provides body-scale operation of battery-free NFC sensors placed arbitrarily around the body.
Unlike previous NFC-enabled textiles with $5-10\%$ narrow coverage of the body, Full-body NFC could cover approximately 70-80\% of the body while achieving sensitive readout of NFC sensors occupying less than 0.3\% of readout coverage.
We showed that Full-body NFC enables both \si{\mW}-class wireless charging and energy-efficient wireless readout for battery-free NFC sensor tags while being robust against user motions.
Full-body NFC can represent a significant step toward--\textit{Internet of Textiles}-- where textiles are not only passive items but active devices in the digital ecosystem.

The current Full-body NFC is still a proof of concept, and has open challenge to discuss.\\
\textbf{Misalignment of Sensor Coil:} the misalignment of the sensor coil with a few centimeter causes the decrease of the power transfer efficiency and increase of the BER, as described in \S~\ref{sec:wireless_charging_capability} and \S~\ref{sec:wireless_communication_capability}.
An array of multiple sensor coils, which covers different coupling position with the reader coil, could ensure that at least some coils maintain optimal alignment with the reader coil. 
\\
\textbf{Data Collision for Multiple NFC Sensor Tags:} the maximum available number of NFC tags is limited to four or five, as described in \S~\ref{sec:application} due to packet collisions.
This limitation is because the NFC protocol in use (the Aloha protocol defined by ISO 15639) is designed to handle only about two to three tags effectively. 
Currently, Full-body NFC can read out four to five tags by reducing the data rate to hundreds of \si{\Hz}. 
Introducing a collision-resilient protocol could allow for tens of NFC sensor tags~\cite{sangar_wichronos_2020,xie_enabling_2024,wang_rf-sifter_2023}.\\
\textbf{Extension of Biomonitoring:} There are several research efforts and developments in non-contact vital sign monitoring using mmWave or UWB-based RF technology under severe movements~\cite{chen_movi-fi_2021,chang_robust_2024,chang_msense_2024,zheng_more-fi_2021,lin_fine-grained_2024}. 
Combining such a high-frequency-based non-contact biosensing with our NFC-based direct biosensing can enable a comprehensive biomonitoring system that leverages the strengths of both approaches. 
The use of NFC-enabled implantable wireless sensors through Full-body NFC can further enhance its capabilities by providing data from within the body~\cite{zhang_nfcapsule_2023}.

\begin{acks}
This work was supported by JST ACT-X JPMJAX21K9, JSPS KAKEN 22K21343, and JST ASPIRE JPMJAP2401. 
\end{acks}

\bibliographystyle{ACM-Reference-Format}
\bibliography{references, reference_other}


\begin{thebibliography}{64}


\ifx \showCODEN    \undefined \def \showCODEN     #1{\unskip}     \fi
\ifx \showDOI      \undefined \def \showDOI       #1{#1}\fi
\ifx \showISBNx    \undefined \def \showISBNx     #1{\unskip}     \fi
\ifx \showISBNxiii \undefined \def \showISBNxiii  #1{\unskip}     \fi
\ifx \showISSN     \undefined \def \showISSN      #1{\unskip}     \fi
\ifx \showLCCN     \undefined \def \showLCCN      #1{\unskip}     \fi
\ifx \shownote     \undefined \def \shownote      #1{#1}          \fi
\ifx \showarticletitle \undefined \def \showarticletitle #1{#1}   \fi
\ifx \showURL      \undefined \def \showURL       {\relax}        \fi
\providecommand\bibfield[2]{#2}
\providecommand\bibinfo[2]{#2}
\providecommand\natexlab[1]{#1}
\providecommand\showeprint[2][]{arXiv:#2}

\bibitem[noa(2024)]%
        {noauthor_high-performance_2024}
 \bibinfo{year}{2024}\natexlab{}.
\newblock \bibinfo{title}{High-performance {ISO}/{IEC} 14443 {A}/{B} frontend {MFRC631} and {MFRC631} plus}.
\newblock
\newblock
\urldef\tempurl%
\url{https://www.nxp.com/docs/en/data-sheet/MFRC631.pdf}
\showURL{%
\tempurl}


\bibitem[Akbar et~al\mbox{.}(2023)]%
        {akbar_underwater_2023}
\bibfield{author}{\bibinfo{person}{Waleed Akbar}, \bibinfo{person}{Ahmed Allam}, {and} \bibinfo{person}{Fadel Adib}.} \bibinfo{year}{2023}\natexlab{}.
\newblock \showarticletitle{The {Underwater} {Backscatter} {Channel}: {Theory}, {Link} {Budget}, and {Experimental} {Validation}}. In \bibinfo{booktitle}{\emph{Proceedings of the 29th {Annual} {International} {Conference} on {Mobile} {Computing} and {Networking}}} \emph{(\bibinfo{series}{{ACM} {MobiCom} '23})}. \bibinfo{publisher}{Association for Computing Machinery}, \bibinfo{address}{New York, NY, USA}, \bibinfo{pages}{1--15}.
\newblock
\showISBNx{978-1-4503-9990-6}
\urldef\tempurl%
\url{https://doi.org/10.1145/3570361.3613265}
\showDOI{\tempurl}


\bibitem[Albaugh et~al\mbox{.}(2019)]%
        {albaugh_digital_2019}
\bibfield{author}{\bibinfo{person}{Lea Albaugh}, \bibinfo{person}{Scott Hudson}, {and} \bibinfo{person}{Lining Yao}.} \bibinfo{year}{2019}\natexlab{}.
\newblock \showarticletitle{Digital {Fabrication} of {Soft} {Actuated} {Objects} by {Machine} {Knitting}}. In \bibinfo{booktitle}{\emph{Proceedings of the 2019 {CHI} {Conference} on {Human} {Factors} in {Computing} {Systems}}} \emph{(\bibinfo{series}{{CHI} '19})}. \bibinfo{publisher}{Association for Computing Machinery}, \bibinfo{address}{New York, NY, USA}, \bibinfo{pages}{1--13}.
\newblock
\showISBNx{978-1-4503-5970-2}
\urldef\tempurl%
\url{https://doi.org/10.1145/3290605.3300414}
\showDOI{\tempurl}


\bibitem[An et~al\mbox{.}(2021)]%
        {an_one_2021}
\bibfield{author}{\bibinfo{person}{Zhenlin An}, \bibinfo{person}{Qiongzheng Lin}, \bibinfo{person}{Xiaopeng Zhao}, \bibinfo{person}{Lei Yang}, \bibinfo{person}{Dongliang Zheng}, \bibinfo{person}{Guiqing Wu}, {and} \bibinfo{person}{Shan Chang}.} \bibinfo{year}{2021}\natexlab{}.
\newblock \showarticletitle{One tag, two codes: identifying optical barcodes with {NFC}}. In \bibinfo{booktitle}{\emph{Proceedings of the 27th {Annual} {International} {Conference} on {Mobile} {Computing} and {Networking}}} \emph{(\bibinfo{series}{{MobiCom} '21})}. \bibinfo{publisher}{Association for Computing Machinery}, \bibinfo{address}{New York, NY, USA}, \bibinfo{pages}{108--120}.
\newblock
\showISBNx{978-1-4503-8342-4}
\urldef\tempurl%
\url{https://doi.org/10.1145/3447993.3448631}
\showDOI{\tempurl}


\bibitem[Bouhassoune et~al\mbox{.}(2019)]%
        {bouhassoune_wireless_2019}
\bibfield{author}{\bibinfo{person}{Ibtissame Bouhassoune}, \bibinfo{person}{Rachid Saadane}, {and} \bibinfo{person}{Abdellah Chehri}.} \bibinfo{year}{2019}\natexlab{}.
\newblock \showarticletitle{Wireless {Body} {Area} {Network} {Based} on {RFID} {System} for {Healthcare} {Monitoring}: {Progress} and {Architectures}}. In \bibinfo{booktitle}{\emph{2019 15th {International} {Conference} on {Signal}-{Image} {Technology} \& {Internet}-{Based} {Systems} ({SITIS})}}. \bibinfo{pages}{416--421}.
\newblock
\urldef\tempurl%
\url{https://doi.org/10.1109/SITIS.2019.00073}
\showDOI{\tempurl}


\bibitem[Chang et~al\mbox{.}(2024a)]%
        {chang_robust_2024}
\bibfield{author}{\bibinfo{person}{Zhaoxin Chang}, \bibinfo{person}{Xinyu Xue}, \bibinfo{person}{Fusang Zhang}, \bibinfo{person}{Jie Xiong}, \bibinfo{person}{Weiyan Chen}, \bibinfo{person}{Badii Jouaber}, {and} \bibinfo{person}{Daqing Zhang}.} \bibinfo{year}{2024}\natexlab{a}.
\newblock \showarticletitle{Robust {Respiration} {Monitoring} {Under} {Body} {Motion} {Interference}}. In \bibinfo{booktitle}{\emph{Proceedings of the 30th {Annual} {International} {Conference} on {Mobile} {Computing} and {Networking}}} \emph{(\bibinfo{series}{{ACM} {MobiCom} '24})}. \bibinfo{publisher}{Association for Computing Machinery}, \bibinfo{address}{New York, NY, USA}, \bibinfo{pages}{1713--1715}.
\newblock
\showISBNx{9798400704895}
\urldef\tempurl%
\url{https://doi.org/10.1145/3636534.3698835}
\showDOI{\tempurl}


\bibitem[Chang et~al\mbox{.}(2024b)]%
        {chang_msense_2024}
\bibfield{author}{\bibinfo{person}{Zhaoxin Chang}, \bibinfo{person}{Fusang Zhang}, \bibinfo{person}{Jie Xiong}, \bibinfo{person}{Weiyan Chen}, {and} \bibinfo{person}{Daqing Zhang}.} \bibinfo{year}{2024}\natexlab{b}.
\newblock \showarticletitle{{MSense}: {Boosting} {Wireless} {Sensing} {Capability} {Under} {Motion} {Interference}}. In \bibinfo{booktitle}{\emph{Proceedings of the 30th {Annual} {International} {Conference} on {Mobile} {Computing} and {Networking}}} \emph{(\bibinfo{series}{{ACM} {MobiCom} '24})}. \bibinfo{publisher}{Association for Computing Machinery}, \bibinfo{address}{New York, NY, USA}, \bibinfo{pages}{108--123}.
\newblock
\showISBNx{9798400704895}
\urldef\tempurl%
\url{https://doi.org/10.1145/3636534.3649350}
\showDOI{\tempurl}


\bibitem[Chen et~al\mbox{.}(2021)]%
        {chen_movi-fi_2021}
\bibfield{author}{\bibinfo{person}{Zhe Chen}, \bibinfo{person}{Tianyue Zheng}, \bibinfo{person}{Chao Cai}, {and} \bibinfo{person}{Jun Luo}.} \bibinfo{year}{2021}\natexlab{}.
\newblock \showarticletitle{{MoVi}-{Fi}: motion-robust vital signs waveform recovery via deep interpreted {RF} sensing}. In \bibinfo{booktitle}{\emph{Proceedings of the 27th {Annual} {International} {Conference} on {Mobile} {Computing} and {Networking}}} \emph{(\bibinfo{series}{{MobiCom} '21})}. \bibinfo{publisher}{Association for Computing Machinery}, \bibinfo{address}{New York, NY, USA}, \bibinfo{pages}{392--405}.
\newblock
\showISBNx{978-1-4503-8342-4}
\urldef\tempurl%
\url{https://doi.org/10.1145/3447993.3483251}
\showDOI{\tempurl}


\bibitem[Cook and Lowe(1982)]%
        {cook_large-inductance_1982}
\bibfield{author}{\bibinfo{person}{Bruce Cook} {and} \bibinfo{person}{I.~J Lowe}.} \bibinfo{year}{1982}\natexlab{}.
\newblock \showarticletitle{A large-inductance, high-frequency, high-\textit{{Q}}, series-tuned coil for {NMR}}.
\newblock \bibinfo{journal}{\emph{Journal of Magnetic Resonance (1969)}} \bibinfo{volume}{49}, \bibinfo{number}{2} (\bibinfo{date}{Sept.} \bibinfo{year}{1982}), \bibinfo{pages}{346--349}.
\newblock
\showISSN{0022-2364}
\urldef\tempurl%
\url{https://doi.org/10.1016/0022-2364(82)90200-1}
\showDOI{\tempurl}


\bibitem[Derogarian~Miyandoab et~al\mbox{.}(2020)]%
        {derogarian_miyandoab_multifunctional_2020}
\bibfield{author}{\bibinfo{person}{Fardin Derogarian~Miyandoab}, \bibinfo{person}{João Canas~Ferreira}, \bibinfo{person}{Vítor~M. Grade~Tavares}, \bibinfo{person}{José Machado~da Silva}, {and} \bibinfo{person}{Fernando~J. Velez}.} \bibinfo{year}{2020}\natexlab{}.
\newblock \showarticletitle{A {Multifunctional} {Integrated} {Circuit} {Router} for {Body} {Area} {Network} {Wearable} {Systems}}.
\newblock \bibinfo{journal}{\emph{IEEE/ACM Transactions on Networking}} \bibinfo{volume}{28}, \bibinfo{number}{5} (\bibinfo{date}{Oct.} \bibinfo{year}{2020}), \bibinfo{pages}{1981--1994}.
\newblock
\showISSN{1558-2566}
\urldef\tempurl%
\url{https://doi.org/10.1109/TNET.2020.3004550}
\showDOI{\tempurl}
\newblock
\shownote{Conference Name: IEEE/ACM Transactions on Networking}.


\bibitem[Gravina and Fortino(2021)]%
        {gravina_wearable_2021}
\bibfield{author}{\bibinfo{person}{Raffaele Gravina} {and} \bibinfo{person}{Giancarlo Fortino}.} \bibinfo{year}{2021}\natexlab{}.
\newblock \showarticletitle{Wearable {Body} {Sensor} {Networks}: {State}-of-the-{Art} and {Research} {Directions}}.
\newblock \bibinfo{journal}{\emph{IEEE Sensors Journal}} \bibinfo{volume}{21}, \bibinfo{number}{11} (\bibinfo{date}{June} \bibinfo{year}{2021}), \bibinfo{pages}{12511--12522}.
\newblock
\showISSN{1558-1748}
\urldef\tempurl%
\url{https://doi.org/10.1109/JSEN.2020.3044447}
\showDOI{\tempurl}
\newblock
\shownote{Conference Name: IEEE Sensors Journal}.


\bibitem[Hajiaghajani et~al\mbox{.}(2021)]%
        {hajiaghajani_textile-integrated_2021}
\bibfield{author}{\bibinfo{person}{Amirhossein Hajiaghajani}, \bibinfo{person}{Amir~Hosein Afandizadeh~Zargari}, \bibinfo{person}{Manik Dautta}, \bibinfo{person}{Abel Jimenez}, \bibinfo{person}{Fadi Kurdahi}, {and} \bibinfo{person}{Peter Tseng}.} \bibinfo{year}{2021}\natexlab{}.
\newblock \showarticletitle{Textile-integrated metamaterials for near-field multibody area networks}.
\newblock \bibinfo{journal}{\emph{Nature Electronics}} \bibinfo{volume}{4}, \bibinfo{number}{11} (\bibinfo{date}{Nov.} \bibinfo{year}{2021}), \bibinfo{pages}{808--817}.
\newblock
\showISSN{2520-1131}
\urldef\tempurl%
\url{https://doi.org/10.1038/s41928-021-00663-0}
\showDOI{\tempurl}
\newblock
\shownote{Publisher: Nature Publishing Group}.


\bibitem[Hu et~al\mbox{.}(2022)]%
        {hu_biotag_2022}
\bibfield{author}{\bibinfo{person}{Bin Hu}, \bibinfo{person}{Tianming Zhao}, \bibinfo{person}{Yan Wang}, \bibinfo{person}{Jerry Cheng}, \bibinfo{person}{Richard Howard}, \bibinfo{person}{Yingying Chen}, {and} \bibinfo{person}{Hao Wan}.} \bibinfo{year}{2022}\natexlab{}.
\newblock \showarticletitle{{BioTag}: robust {RFID}-based continuous user verification using physiological features from respiration}. In \bibinfo{booktitle}{\emph{Proceedings of the {Twenty}-{Third} {International} {Symposium} on {Theory}, {Algorithmic} {Foundations}, and {Protocol} {Design} for {Mobile} {Networks} and {Mobile} {Computing}}} \emph{(\bibinfo{series}{{MobiHoc} '22})}. \bibinfo{publisher}{Association for Computing Machinery}, \bibinfo{address}{New York, NY, USA}, \bibinfo{pages}{191--200}.
\newblock
\showISBNx{978-1-4503-9165-8}
\urldef\tempurl%
\url{https://doi.org/10.1145/3492866.3549718}
\showDOI{\tempurl}


\bibitem[Iyer et~al\mbox{.}(2020)]%
        {iyer_airdropping_2020}
\bibfield{author}{\bibinfo{person}{Vikram Iyer}, \bibinfo{person}{Maruchi Kim}, \bibinfo{person}{Shirley Xue}, \bibinfo{person}{Anran Wang}, {and} \bibinfo{person}{Shyamnath Gollakota}.} \bibinfo{year}{2020}\natexlab{}.
\newblock \showarticletitle{Airdropping sensor networks from drones and insects}. In \bibinfo{booktitle}{\emph{Proceedings of the 26th {Annual} {International} {Conference} on {Mobile} {Computing} and {Networking}}} \emph{(\bibinfo{series}{{MobiCom} '20})}. \bibinfo{publisher}{Association for Computing Machinery}, \bibinfo{address}{New York, NY, USA}, \bibinfo{pages}{1--14}.
\newblock
\showISBNx{978-1-4503-7085-1}
\urldef\tempurl%
\url{https://doi.org/10.1145/3372224.3419981}
\showDOI{\tempurl}


\bibitem[Iyer et~al\mbox{.}(2019)]%
        {iyer_living_2019}
\bibfield{author}{\bibinfo{person}{Vikram Iyer}, \bibinfo{person}{Rajalakshmi Nandakumar}, \bibinfo{person}{Anran Wang}, \bibinfo{person}{Sawyer~B. Fuller}, {and} \bibinfo{person}{Shyamnath Gollakota}.} \bibinfo{year}{2019}\natexlab{}.
\newblock \showarticletitle{Living {IoT}: {A} {Flying} {Wireless} {Platform} on {Live} {Insects}}. In \bibinfo{booktitle}{\emph{The 25th {Annual} {International} {Conference} on {Mobile} {Computing} and {Networking}}} \emph{(\bibinfo{series}{{MobiCom} '19})}. \bibinfo{publisher}{Association for Computing Machinery}, \bibinfo{address}{New York, NY, USA}, \bibinfo{pages}{1--15}.
\newblock
\showISBNx{978-1-4503-6169-9}
\urldef\tempurl%
\url{https://doi.org/10.1145/3300061.3300136}
\showDOI{\tempurl}


\bibitem[Jiang et~al\mbox{.}(2020)]%
        {jiang_smart_2020}
\bibfield{author}{\bibinfo{person}{Yutong Jiang}, \bibinfo{person}{Kewen Pan}, \bibinfo{person}{Ting Leng}, {and} \bibinfo{person}{Zhirun Hu}.} \bibinfo{year}{2020}\natexlab{}.
\newblock \showarticletitle{Smart {Textile} {Integrated} {Wireless} {Powered} {Near} {Field} {Communication} {Body} {Temperature} and {Sweat} {Sensing} {System}}.
\newblock \bibinfo{journal}{\emph{IEEE Journal of Electromagnetics, RF and Microwaves in Medicine and Biology}} \bibinfo{volume}{4}, \bibinfo{number}{3} (\bibinfo{date}{Sept.} \bibinfo{year}{2020}), \bibinfo{pages}{164--170}.
\newblock
\showISSN{2469-7257}
\urldef\tempurl%
\url{https://doi.org/10.1109/JERM.2019.2929676}
\showDOI{\tempurl}
\newblock
\shownote{Conference Name: IEEE Journal of Electromagnetics, RF and Microwaves in Medicine and Biology}.


\bibitem[Jiao et~al\mbox{.}(2024)]%
        {jiao_zeroecg_2024}
\bibfield{author}{\bibinfo{person}{Wenli Jiao}, \bibinfo{person}{Ju Wang}, \bibinfo{person}{Xinzhuo Gao}, \bibinfo{person}{Long Du}, \bibinfo{person}{Yanlin Li}, \bibinfo{person}{Lili Zhao}, \bibinfo{person}{Dingyi Fang}, {and} \bibinfo{person}{Xiaojiang Chen}.} \bibinfo{year}{2024}\natexlab{}.
\newblock \showarticletitle{{ZEROECG}: {Zero}-{Sensation} {ECG} {Monitoring} {By} {Exploring} {RFID} {MOSFET}}. In \bibinfo{booktitle}{\emph{Proceedings of the 30th {Annual} {International} {Conference} on {Mobile} {Computing} and {Networking}}} \emph{(\bibinfo{series}{{ACM} {MobiCom} '24})}. \bibinfo{publisher}{Association for Computing Machinery}, \bibinfo{address}{New York, NY, USA}, \bibinfo{pages}{1237--1251}.
\newblock
\showISBNx{9798400704895}
\urldef\tempurl%
\url{https://doi.org/10.1145/3636534.3690690}
\showDOI{\tempurl}


\bibitem[Kanada et~al\mbox{.}(2025)]%
        {kanada_joint-repositionable_2024}
\bibfield{author}{\bibinfo{person}{Ayato Kanada}, \bibinfo{person}{Ryo Takahashi}, \bibinfo{person}{Keito Hayashi}, \bibinfo{person}{Ryusuke Hosaka}, \bibinfo{person}{Wakako Yukita}, \bibinfo{person}{Yasutaka Nakashima}, \bibinfo{person}{Tomoyuki Yokota}, \bibinfo{person}{Takao Someya}, \bibinfo{person}{Mitsuhiro Kamezaki}, \bibinfo{person}{Yoshihiro Kawahara}, {and} \bibinfo{person}{Motoji Yamamoto}.} \bibinfo{year}{2025}\natexlab{}.
\newblock \bibinfo{title}{Joint-repositionable {Inner}-wireless {Planar} {Snake} {Robot}}.
\newblock
\newblock


\bibitem[Kim et~al\mbox{.}(2024)]%
        {kim_strain-invariant_2024}
\bibfield{author}{\bibinfo{person}{Sun~Hong Kim}, \bibinfo{person}{Abdul Basir}, \bibinfo{person}{Raudel Avila}, \bibinfo{person}{Jaeman Lim}, \bibinfo{person}{Seong~Woo Hong}, \bibinfo{person}{Geonoh Choe}, \bibinfo{person}{Joo~Hwan Shin}, \bibinfo{person}{Jin~Hee Hwang}, \bibinfo{person}{Sun~Young Park}, \bibinfo{person}{Jiho Joo}, \bibinfo{person}{Chanmi Lee}, \bibinfo{person}{Jaehoon Choi}, \bibinfo{person}{Byunghun Lee}, \bibinfo{person}{Kwang-Seong Choi}, \bibinfo{person}{Sungmook Jung}, \bibinfo{person}{Tae-il Kim}, \bibinfo{person}{Hyoungsuk Yoo}, {and} \bibinfo{person}{Yei~Hwan Jung}.} \bibinfo{year}{2024}\natexlab{}.
\newblock \showarticletitle{Strain-invariant stretchable radio-frequency electronics}.
\newblock \bibinfo{journal}{\emph{Nature}} \bibinfo{volume}{629}, \bibinfo{number}{8014} (\bibinfo{date}{May} \bibinfo{year}{2024}), \bibinfo{pages}{1047--1054}.
\newblock
\showISSN{1476-4687}
\urldef\tempurl%
\url{https://doi.org/10.1038/s41586-024-07383-3}
\showDOI{\tempurl}
\newblock
\shownote{Publisher: Nature Publishing Group}.


\bibitem[Kong et~al\mbox{.}(2024)]%
        {kong_power-over-skin_2024}
\bibfield{author}{\bibinfo{person}{Andy Kong}, \bibinfo{person}{Daehwa Kim}, {and} \bibinfo{person}{Chris Harrison}.} \bibinfo{year}{2024}\natexlab{}.
\newblock \showarticletitle{Power-over-{Skin}: {Full}-{Body} {Wearables} {Powered} {By} {Intra}-{Body} {RF} {Energy}}. In \bibinfo{booktitle}{\emph{Proceedings of the 37th {Annual} {ACM} {Symposium} on {User} {Interface} {Software} and {Technology}}}. \bibinfo{publisher}{ACM}, \bibinfo{address}{Pittsburgh PA USA}, \bibinfo{pages}{1--13}.
\newblock
\showISBNx{9798400706288}
\urldef\tempurl%
\url{https://doi.org/10.1145/3654777.3676394}
\showDOI{\tempurl}


\bibitem[Li et~al\mbox{.}(2021)]%
        {li_body-coupled_2021}
\bibfield{author}{\bibinfo{person}{Jiamin Li}, \bibinfo{person}{Yilong Dong}, \bibinfo{person}{Jeong~Hoan Park}, {and} \bibinfo{person}{Jerald Yoo}.} \bibinfo{year}{2021}\natexlab{}.
\newblock \showarticletitle{Body-coupled power transmission and energy harvesting}.
\newblock \bibinfo{journal}{\emph{Nature Electronics}} \bibinfo{volume}{4}, \bibinfo{number}{7} (\bibinfo{date}{July} \bibinfo{year}{2021}), \bibinfo{pages}{530--538}.
\newblock
\showISSN{2520-1131}
\urldef\tempurl%
\url{https://doi.org/10.1038/s41928-021-00592-y}
\showDOI{\tempurl}
\newblock
\shownote{Publisher: Nature Publishing Group}.


\bibitem[Li et~al\mbox{.}(2020)]%
        {li_internet--microchips_2020}
\bibfield{author}{\bibinfo{person}{Songfan Li}, \bibinfo{person}{Chong Zhang}, \bibinfo{person}{Yihang Song}, \bibinfo{person}{Hui Zheng}, \bibinfo{person}{Lu Liu}, \bibinfo{person}{Li Lu}, {and} \bibinfo{person}{Mo Li}.} \bibinfo{year}{2020}\natexlab{}.
\newblock \showarticletitle{Internet-of-microchips: direct radio-to-bus communication with {SPI} backscatter}. In \bibinfo{booktitle}{\emph{Proceedings of the 26th {Annual} {International} {Conference} on {Mobile} {Computing} and {Networking}}} \emph{(\bibinfo{series}{{MobiCom} '20})}. \bibinfo{publisher}{Association for Computing Machinery}, \bibinfo{address}{New York, NY, USA}, \bibinfo{pages}{1--14}.
\newblock
\showISBNx{978-1-4503-7085-1}
\urldef\tempurl%
\url{https://doi.org/10.1145/3372224.3419182}
\showDOI{\tempurl}


\bibitem[Li et~al\mbox{.}(2025)]%
        {li_plug-n-play_2025}
\bibfield{author}{\bibinfo{person}{Yifan Li}, \bibinfo{person}{Ryo Takahashi}, \bibinfo{person}{Wakako Yukita}, \bibinfo{person}{Kanata Matsutani}, \bibinfo{person}{Cedric Caremel}, \bibinfo{person}{Yuhiro Iwamoto}, \bibinfo{person}{Sunghoon Lee}, \bibinfo{person}{Tomoyuki Yokota}, \bibinfo{person}{Takao Someya}, {and} \bibinfo{person}{Yoshihiro Kawahara}.} \bibinfo{year}{2025}\natexlab{}.
\newblock \showarticletitle{Plug-n-play e-knit: prototyping large-area e-textiles using machine-knitted magnetically-repositionable sensor networks}. In \bibinfo{booktitle}{\emph{Proceedings of the {Nineteenth} {International} {Conference} on {Tangible}, {Embedded}, and {Embodied} {Interaction}}} \emph{(\bibinfo{series}{{TEI} '25})}. \bibinfo{publisher}{Association for Computing Machinery}, \bibinfo{address}{New York, NY, USA}, \bibinfo{pages}{1--7}.
\newblock
\showISBNx{9798400711978}
\urldef\tempurl%
\url{https://doi.org/10.1145/3689050.3705973}
\showDOI{\tempurl}


\bibitem[Libanori et~al\mbox{.}(2022)]%
        {libanori_smart_2022}
\bibfield{author}{\bibinfo{person}{Alberto Libanori}, \bibinfo{person}{Guorui Chen}, \bibinfo{person}{Xun Zhao}, \bibinfo{person}{Yihao Zhou}, {and} \bibinfo{person}{Jun Chen}.} \bibinfo{year}{2022}\natexlab{}.
\newblock \showarticletitle{Smart textiles for personalized healthcare}.
\newblock \bibinfo{journal}{\emph{Nature Electronics}} \bibinfo{volume}{5}, \bibinfo{number}{3} (\bibinfo{date}{March} \bibinfo{year}{2022}), \bibinfo{pages}{142--156}.
\newblock
\showISSN{2520-1131}
\urldef\tempurl%
\url{https://doi.org/10.1038/s41928-022-00723-z}
\showDOI{\tempurl}
\newblock
\shownote{Publisher: Nature Publishing Group}.


\bibitem[Lin et~al\mbox{.}(2024)]%
        {lin_fine-grained_2024}
\bibfield{author}{\bibinfo{person}{Chi Lin}, \bibinfo{person}{Zhaohe Wang}, \bibinfo{person}{Jie Xiong}, \bibinfo{person}{Fengqi Li}, {and} \bibinfo{person}{Guowei Wu}.} \bibinfo{year}{2024}\natexlab{}.
\newblock \showarticletitle{Fine-grained {Textile} {Moisture} {Sensing} with {Commodity} {UWB}}. In \bibinfo{booktitle}{\emph{Proceedings of the 30th {Annual} {International} {Conference} on {Mobile} {Computing} and {Networking}}} \emph{(\bibinfo{series}{{ACM} {MobiCom} '24})}. \bibinfo{publisher}{Association for Computing Machinery}, \bibinfo{address}{New York, NY, USA}, \bibinfo{pages}{1074--1088}.
\newblock
\showISBNx{9798400704895}
\urldef\tempurl%
\url{https://doi.org/10.1145/3636534.3690679}
\showDOI{\tempurl}


\bibitem[Lin et~al\mbox{.}(2020)]%
        {lin_wireless_2020}
\bibfield{author}{\bibinfo{person}{Rongzhou Lin}, \bibinfo{person}{Han-Joon Kim}, \bibinfo{person}{Sippanat Achavananthadith}, \bibinfo{person}{Selman~A. Kurt}, \bibinfo{person}{Shawn C.~C. Tan}, \bibinfo{person}{Haicheng Yao}, \bibinfo{person}{Benjamin C.~K. Tee}, \bibinfo{person}{Jason K.~W. Lee}, {and} \bibinfo{person}{John~S. Ho}.} \bibinfo{year}{2020}\natexlab{}.
\newblock \showarticletitle{Wireless battery-free body sensor networks using near-field-enabled clothing}.
\newblock \bibinfo{journal}{\emph{Nature Communications}} \bibinfo{volume}{11}, \bibinfo{number}{1} (\bibinfo{date}{Jan.} \bibinfo{year}{2020}), \bibinfo{pages}{444}.
\newblock
\showISSN{2041-1723}
\urldef\tempurl%
\url{https://doi.org/10.1038/s41467-020-14311-2}
\showDOI{\tempurl}
\newblock
\shownote{Publisher: Nature Publishing Group}.


\bibitem[Lin et~al\mbox{.}(2022)]%
        {lin_digitally-embroidered_2022}
\bibfield{author}{\bibinfo{person}{Rongzhou Lin}, \bibinfo{person}{Han-Joon Kim}, \bibinfo{person}{Sippanat Achavananthadith}, \bibinfo{person}{Ze Xiong}, \bibinfo{person}{Jason K.~W. Lee}, \bibinfo{person}{Yong~Lin Kong}, {and} \bibinfo{person}{John~S. Ho}.} \bibinfo{year}{2022}\natexlab{}.
\newblock \showarticletitle{Digitally-embroidered liquid metal electronic textiles for wearable wireless systems}.
\newblock \bibinfo{journal}{\emph{Nature Communications}} \bibinfo{volume}{13}, \bibinfo{number}{1} (\bibinfo{date}{April} \bibinfo{year}{2022}), \bibinfo{pages}{2190}.
\newblock
\showISSN{2041-1723}
\urldef\tempurl%
\url{https://doi.org/10.1038/s41467-022-29859-4}
\showDOI{\tempurl}
\newblock
\shownote{Publisher: Nature Publishing Group}.


\bibitem[Natarajan et~al\mbox{.}(2009)]%
        {natarajan_link_2009}
\bibfield{author}{\bibinfo{person}{Anirudh Natarajan}, \bibinfo{person}{Buddhika de Silva}, \bibinfo{person}{Kok-Kiong Yap}, {and} \bibinfo{person}{Mehul Motani}.} \bibinfo{year}{2009}\natexlab{}.
\newblock \showarticletitle{Link layer behavior of body area networks at 2.4 {GHz}}. In \bibinfo{booktitle}{\emph{Proceedings of the 15th annual international conference on {Mobile} computing and networking}} \emph{(\bibinfo{series}{{MobiCom} '09})}. \bibinfo{publisher}{Association for Computing Machinery}, \bibinfo{address}{New York, NY, USA}, \bibinfo{pages}{241--252}.
\newblock
\showISBNx{978-1-60558-702-8}
\urldef\tempurl%
\url{https://doi.org/10.1145/1614320.1614347}
\showDOI{\tempurl}


\bibitem[Negra et~al\mbox{.}(2016)]%
        {negra_wireless_2016}
\bibfield{author}{\bibinfo{person}{Rim Negra}, \bibinfo{person}{Imen Jemili}, {and} \bibinfo{person}{Abdelfettah Belghith}.} \bibinfo{year}{2016}\natexlab{}.
\newblock \showarticletitle{Wireless {Body} {Area} {Networks}: {Applications} and {Technologies}}.
\newblock \bibinfo{journal}{\emph{Procedia Computer Science}}  \bibinfo{volume}{83} (\bibinfo{date}{Jan.} \bibinfo{year}{2016}), \bibinfo{pages}{1274--1281}.
\newblock
\showISSN{1877-0509}
\urldef\tempurl%
\url{https://doi.org/10.1016/j.procs.2016.04.266}
\showDOI{\tempurl}


\bibitem[Niu et~al\mbox{.}(2019)]%
        {niu_wireless_2019}
\bibfield{author}{\bibinfo{person}{Simiao Niu}, \bibinfo{person}{Naoji Matsuhisa}, \bibinfo{person}{Levent Beker}, \bibinfo{person}{Jinxing Li}, \bibinfo{person}{Sihong Wang}, \bibinfo{person}{Jiechen Wang}, \bibinfo{person}{Yuanwen Jiang}, \bibinfo{person}{Xuzhou Yan}, \bibinfo{person}{Youngjun Yun}, \bibinfo{person}{William Burnett}, \bibinfo{person}{Ada S.~Y. Poon}, \bibinfo{person}{Jeffery B.-H. Tok}, \bibinfo{person}{Xiaodong Chen}, {and} \bibinfo{person}{Zhenan Bao}.} \bibinfo{year}{2019}\natexlab{}.
\newblock \showarticletitle{A wireless body area sensor network based on stretchable passive tags}.
\newblock \bibinfo{journal}{\emph{Nature Electronics}} \bibinfo{volume}{2}, \bibinfo{number}{8} (\bibinfo{date}{Aug.} \bibinfo{year}{2019}), \bibinfo{pages}{361--368}.
\newblock
\showISSN{2520-1131}
\urldef\tempurl%
\url{https://doi.org/10.1038/s41928-019-0286-2}
\showDOI{\tempurl}
\newblock
\shownote{Publisher: Nature Publishing Group}.


\bibitem[Noda(2019)]%
        {noda_wearable_2019}
\bibfield{author}{\bibinfo{person}{Akihito Noda}.} \bibinfo{year}{2019}\natexlab{}.
\newblock \showarticletitle{Wearable {NFC} {Reader} and {Sensor} {Tag} for {Health} {Monitoring}}. In \bibinfo{booktitle}{\emph{2019 {IEEE} {Biomedical} {Circuits} and {Systems} {Conference} ({BioCAS})}}. \bibinfo{pages}{1--4}.
\newblock
\urldef\tempurl%
\url{https://doi.org/10.1109/BIOCAS.2019.8919200}
\showDOI{\tempurl}
\newblock
\shownote{ISSN: 2163-4025}.


\bibitem[Noda and Shinoda(2019)]%
        {Noda2019Inter-ICTextile}
\bibfield{author}{\bibinfo{person}{Akihito Noda} {and} \bibinfo{person}{Hiroyuki Shinoda}.} \bibinfo{year}{2019}\natexlab{}.
\newblock \showarticletitle{{Inter-IC for Wearables (I2We): Power and Data Transfer over Double-Sided Conductive Textile}}.
\newblock \bibinfo{journal}{\emph{IEEE Transactions on Biomedical Circuits and Systems}} \bibinfo{volume}{13}, \bibinfo{number}{1} (\bibinfo{year}{2019}), \bibinfo{pages}{80--90}.
\newblock


\bibitem[Poupyrev et~al\mbox{.}(2016)]%
        {poupyrev_project_2016}
\bibfield{author}{\bibinfo{person}{Ivan Poupyrev}, \bibinfo{person}{Nan-Wei Gong}, \bibinfo{person}{Shiho Fukuhara}, \bibinfo{person}{Mustafa~Emre Karagozler}, \bibinfo{person}{Carsten Schwesig}, {and} \bibinfo{person}{Karen~E. Robinson}.} \bibinfo{year}{2016}\natexlab{}.
\newblock \showarticletitle{Project {Jacquard}: {Interactive} {Digital} {Textiles} at {Scale}}. In \bibinfo{booktitle}{\emph{Proceedings of the 2016 {CHI} {Conference} on {Human} {Factors} in {Computing} {Systems}}} \emph{(\bibinfo{series}{{CHI} '16})}. \bibinfo{publisher}{Association for Computing Machinery}, \bibinfo{address}{New York, NY, USA}, \bibinfo{pages}{4216--4227}.
\newblock
\showISBNx{978-1-4503-3362-7}
\urldef\tempurl%
\url{https://doi.org/10.1145/2858036.2858176}
\showDOI{\tempurl}


\bibitem[Sangar and Krishnaswamy(2020)]%
        {sangar_wichronos_2020}
\bibfield{author}{\bibinfo{person}{Yaman Sangar} {and} \bibinfo{person}{Bhuvana Krishnaswamy}.} \bibinfo{year}{2020}\natexlab{}.
\newblock \showarticletitle{{WiChronos}: energy-efficient modulation for long-range, large-scale wireless networks}. In \bibinfo{booktitle}{\emph{Proceedings of the 26th {Annual} {International} {Conference} on {Mobile} {Computing} and {Networking}}} \emph{(\bibinfo{series}{{MobiCom} '20})}. \bibinfo{publisher}{Association for Computing Machinery}, \bibinfo{address}{New York, NY, USA}, \bibinfo{pages}{1--14}.
\newblock
\showISBNx{978-1-4503-7085-1}
\urldef\tempurl%
\url{https://doi.org/10.1145/3372224.3380898}
\showDOI{\tempurl}


\bibitem[Sato et~al\mbox{.}(2025)]%
        {sato2025mems}
\bibfield{author}{\bibinfo{person}{Takashi Sato}, \bibinfo{person}{Shinto Watanabe}, \bibinfo{person}{Ryo Takahashi}, \bibinfo{person}{Wakako Yukita}, \bibinfo{person}{Tomoyuki Yokota}, \bibinfo{person}{Takao Someya}, \bibinfo{person}{Yoshihiro Kawahara}, \bibinfo{person}{Eiji Iwase}, {and} \bibinfo{person}{Junya Kurumida}.} \bibinfo{year}{2025}\natexlab{}.
\newblock \showarticletitle{Friction Jointing of Distributed Rigid Capacitors to Stretchable Liquid Metal Coil for Full-body Wireless Charging Clothing}. In \bibinfo{booktitle}{\emph{2025 IEEE 38th International Conference on Micro Electro Mechanical Systems (MEMS)}} (Kaohsiung, Taiwan). \bibinfo{pages}{181–184}.
\newblock


\bibitem[Sato et~al\mbox{.}(2021)]%
        {sato_method_2021}
\bibfield{author}{\bibinfo{person}{Takashi Sato}, \bibinfo{person}{Kento Yamagishi}, \bibinfo{person}{Michinao Hashimoto}, {and} \bibinfo{person}{Eiji Iwase}.} \bibinfo{year}{2021}\natexlab{}.
\newblock \showarticletitle{Method to {Reduce} the {Contact} {Resistivity} between {Galinstan} and a {Copper} {Electrode} for {Electrical} {Connection} in {Flexible} {Devices}}.
\newblock \bibinfo{journal}{\emph{ACS Applied Materials \& Interfaces}} \bibinfo{volume}{13}, \bibinfo{number}{15} (\bibinfo{date}{April} \bibinfo{year}{2021}), \bibinfo{pages}{18247--18254}.
\newblock
\showISSN{1944-8244}
\urldef\tempurl%
\url{https://doi.org/10.1021/acsami.1c00431}
\showDOI{\tempurl}
\newblock
\shownote{Publisher: American Chemical Society}.


\bibitem[Shi et~al\mbox{.}(2021)]%
        {shi_large-area_2021}
\bibfield{author}{\bibinfo{person}{Xiang Shi}, \bibinfo{person}{Yong Zuo}, \bibinfo{person}{Peng Zhai}, \bibinfo{person}{Jiahao Shen}, \bibinfo{person}{Yangyiwei Yang}, \bibinfo{person}{Zhen Gao}, \bibinfo{person}{Meng Liao}, \bibinfo{person}{Jingxia Wu}, \bibinfo{person}{Jiawei Wang}, \bibinfo{person}{Xiaojie Xu}, \bibinfo{person}{Qi Tong}, \bibinfo{person}{Bo Zhang}, \bibinfo{person}{Bingjie Wang}, \bibinfo{person}{Xuemei Sun}, \bibinfo{person}{Lihua Zhang}, \bibinfo{person}{Qibing Pei}, \bibinfo{person}{Dayong Jin}, \bibinfo{person}{Peining Chen}, {and} \bibinfo{person}{Huisheng Peng}.} \bibinfo{year}{2021}\natexlab{}.
\newblock \showarticletitle{Large-area display textiles integrated with functional systems}.
\newblock \bibinfo{journal}{\emph{Nature}} \bibinfo{volume}{591}, \bibinfo{number}{7849} (\bibinfo{date}{March} \bibinfo{year}{2021}), \bibinfo{pages}{240--245}.
\newblock
\showISSN{1476-4687}
\urldef\tempurl%
\url{https://doi.org/10.1038/s41586-021-03295-8}
\showDOI{\tempurl}
\newblock
\shownote{Publisher: Nature Publishing Group}.


\bibitem[Shukla et~al\mbox{.}(2019)]%
        {shukla_skinnypower_2019}
\bibfield{author}{\bibinfo{person}{Rishi Shukla}, \bibinfo{person}{Neev Kiran}, \bibinfo{person}{Rui Wang}, \bibinfo{person}{Jeremy Gummeson}, {and} \bibinfo{person}{Sunghoon~Ivan Lee}.} \bibinfo{year}{2019}\natexlab{}.
\newblock \showarticletitle{{SkinnyPower}: enabling batteryless wearable sensors via intra-body power transfer}. In \bibinfo{booktitle}{\emph{Proceedings of the 17th {Conference} on {Embedded} {Networked} {Sensor} {Systems}}} \emph{(\bibinfo{series}{{SenSys} '19})}. \bibinfo{publisher}{Association for Computing Machinery}, \bibinfo{address}{New York, NY, USA}, \bibinfo{pages}{68--82}.
\newblock
\showISBNx{978-1-4503-6950-3}
\urldef\tempurl%
\url{https://doi.org/10.1145/3356250.3360034}
\showDOI{\tempurl}


\bibitem[Takahashi et~al\mbox{.}(2020)]%
        {takahashi_telemetring_2020}
\bibfield{author}{\bibinfo{person}{Ryo Takahashi}, \bibinfo{person}{Masaaki Fukumoto}, \bibinfo{person}{Changyo Han}, \bibinfo{person}{Takuya Sasatani}, \bibinfo{person}{Yoshiaki Narusue}, {and} \bibinfo{person}{Yoshihiro Kawahara}.} \bibinfo{year}{2020}\natexlab{}.
\newblock \showarticletitle{{TelemetRing}: {A} {Batteryless} and {Wireless} {Ring}-shaped {Keyboard} using {Passive} {Inductive} {Telemetry}}. In \bibinfo{booktitle}{\emph{Proceedings of the 33rd {Annual} {ACM} {Symposium} on {User} {Interface} {Software} and {Technology}}} \emph{(\bibinfo{series}{{UIST} '20})}. \bibinfo{publisher}{Association for Computing Machinery}, \bibinfo{address}{New York, NY, USA}, \bibinfo{pages}{1161--1168}.
\newblock
\showISBNx{978-1-4503-7514-6}
\urldef\tempurl%
\url{https://doi.org/10.1145/3379337.3415873}
\showDOI{\tempurl}


\bibitem[Takahashi et~al\mbox{.}(2024)]%
        {takahashi_picoring_2024}
\bibfield{author}{\bibinfo{person}{Ryo Takahashi}, \bibinfo{person}{Eric Whitmire}, \bibinfo{person}{Roger Boldu}, \bibinfo{person}{Shiu Ng}, \bibinfo{person}{Wolf Kienzle}, {and} \bibinfo{person}{Hrvoje Benko}.} \bibinfo{year}{2024}\natexlab{}.
\newblock \showarticletitle{{picoRing}: battery-free rings for subtle thumb-to-index input}. In \bibinfo{booktitle}{\emph{Proceedings of the 37th {Annual} {ACM} {Symposium} on {User} {Interface} {Software} and {Technology}}} \emph{(\bibinfo{series}{{UIST} '24})}. \bibinfo{publisher}{Association for Computing Machinery}, \bibinfo{address}{New York, NY, USA}, \bibinfo{pages}{1--11}.
\newblock
\showISBNx{9798400706288}
\urldef\tempurl%
\url{https://doi.org/10.1145/3654777.3676365}
\showDOI{\tempurl}


\bibitem[Takahashi et~al\mbox{.}(2022a)]%
        {takahashi_twin_2022}
\bibfield{author}{\bibinfo{person}{Ryo Takahashi}, \bibinfo{person}{Wakako Yukita}, \bibinfo{person}{Takuya Sasatani}, \bibinfo{person}{Tomoyuki Yokota}, \bibinfo{person}{Takao Someya}, {and} \bibinfo{person}{Yoshihiro Kawahara}.} \bibinfo{year}{2022}\natexlab{a}.
\newblock \showarticletitle{Twin {Meander} {Coil}: {Sensitive} {Readout} of {Battery}-free {On}-body {Wireless} {Sensors} {Using} {Body}-scale {Meander} {Coils}}.
\newblock \bibinfo{journal}{\emph{Proc. ACM Interact. Mob. Wearable Ubiquitous Technol.}} \bibinfo{volume}{5}, \bibinfo{number}{4} (\bibinfo{year}{2022}), \bibinfo{pages}{179:1--179:21}.
\newblock
\urldef\tempurl%
\url{https://doi.org/10.1145/3494996}
\showDOI{\tempurl}


\bibitem[Takahashi et~al\mbox{.}(2022b)]%
        {takahashi_meander_2022}
\bibfield{author}{\bibinfo{person}{Ryo Takahashi}, \bibinfo{person}{Wakako Yukita}, \bibinfo{person}{Tomoyuki Yokota}, \bibinfo{person}{Takao Someya}, {and} \bibinfo{person}{Yoshihiro Kawahara}.} \bibinfo{year}{2022}\natexlab{b}.
\newblock \showarticletitle{Meander {Coil}++: {A} {Body}-scale {Wireless} {Power} {Transmission} {Using} {Safe}-to-body and {Energy}-efficient {Transmitter} {Coil}}. In \bibinfo{booktitle}{\emph{{CHI} {Conference} on {Human} {Factors} in {Computing} {Systems}}}. \bibinfo{publisher}{ACM}, \bibinfo{address}{New Orleans LA USA}, \bibinfo{pages}{1--12}.
\newblock
\showISBNx{978-1-4503-9157-3}
\urldef\tempurl%
\url{https://doi.org/10.1145/3491102.3502119}
\showDOI{\tempurl}


\bibitem[Talpur et~al\mbox{.}(2019)]%
        {talpur_validation_2019}
\bibfield{author}{\bibinfo{person}{Anum Talpur}, \bibinfo{person}{Faisal~Karim Shaikh}, \bibinfo{person}{Natasha Baloch}, \bibinfo{person}{Emad Felemban}, \bibinfo{person}{Abdelmajid Khelil}, {and} \bibinfo{person}{Muhammad~Mahtab Alam}.} \bibinfo{year}{2019}\natexlab{}.
\newblock \showarticletitle{Validation of {Wired} and {Wireless} {Interconnected} {Body} {Sensor} {Networks}}.
\newblock \bibinfo{journal}{\emph{Sensors}} \bibinfo{volume}{19}, \bibinfo{number}{17} (\bibinfo{date}{Jan.} \bibinfo{year}{2019}), \bibinfo{pages}{3697}.
\newblock
\showISSN{1424-8220}
\urldef\tempurl%
\url{https://doi.org/10.3390/s19173697}
\showDOI{\tempurl}
\newblock
\shownote{Number: 17 Publisher: Multidisciplinary Digital Publishing Institute}.


\bibitem[Tao et~al\mbox{.}(2023)]%
        {tao_magnetic_2023}
\bibfield{author}{\bibinfo{person}{Bill Tao}, \bibinfo{person}{Emerson Sie}, \bibinfo{person}{Jay Shenoy}, {and} \bibinfo{person}{Deepak Vasisht}.} \bibinfo{year}{2023}\natexlab{}.
\newblock \showarticletitle{Magnetic {Backscatter} for {In}-body {Communication} and {Localization}}. In \bibinfo{booktitle}{\emph{Proceedings of the 29th {Annual} {International} {Conference} on {Mobile} {Computing} and {Networking}}}. \bibinfo{publisher}{ACM}, \bibinfo{address}{Madrid Spain}, \bibinfo{pages}{1--15}.
\newblock
\showISBNx{978-1-4503-9990-6}
\urldef\tempurl%
\url{https://doi.org/10.1145/3570361.3613301}
\showDOI{\tempurl}


\bibitem[Tian et~al\mbox{.}(2019)]%
        {tian_wireless_2019}
\bibfield{author}{\bibinfo{person}{Xi Tian}, \bibinfo{person}{Pui~Mun Lee}, \bibinfo{person}{Yu~Jun Tan}, \bibinfo{person}{Tina L.~Y. Wu}, \bibinfo{person}{Haicheng Yao}, \bibinfo{person}{Mengying Zhang}, \bibinfo{person}{Zhipeng Li}, \bibinfo{person}{Kian~Ann Ng}, \bibinfo{person}{Benjamin C.~K. Tee}, {and} \bibinfo{person}{John~S. Ho}.} \bibinfo{year}{2019}\natexlab{}.
\newblock \showarticletitle{Wireless body sensor networks based on metamaterial textiles}.
\newblock \bibinfo{journal}{\emph{Nature Electronics}} \bibinfo{volume}{2}, \bibinfo{number}{6} (\bibinfo{date}{June} \bibinfo{year}{2019}), \bibinfo{pages}{243--251}.
\newblock
\showISSN{2520-1131}
\urldef\tempurl%
\url{https://doi.org/10.1038/s41928-019-0257-7}
\showDOI{\tempurl}
\newblock
\shownote{Publisher: Nature Publishing Group}.


\bibitem[Tian et~al\mbox{.}(2023)]%
        {tian_implant--implant_2023}
\bibfield{author}{\bibinfo{person}{Xi Tian}, \bibinfo{person}{Qihang Zeng}, \bibinfo{person}{Selman~A. Kurt}, \bibinfo{person}{Renee~R. Li}, \bibinfo{person}{Dat~T. Nguyen}, \bibinfo{person}{Ze Xiong}, \bibinfo{person}{Zhipeng Li}, \bibinfo{person}{Xin Yang}, \bibinfo{person}{Xiao Xiao}, \bibinfo{person}{Changsheng Wu}, \bibinfo{person}{Benjamin C.~K. Tee}, \bibinfo{person}{Denys Nikolayev}, \bibinfo{person}{Christopher~J. Charles}, {and} \bibinfo{person}{John~S. Ho}.} \bibinfo{year}{2023}\natexlab{}.
\newblock \showarticletitle{Implant-to-implant wireless networking with metamaterial textiles}.
\newblock \bibinfo{journal}{\emph{Nature Communications}} \bibinfo{volume}{14}, \bibinfo{number}{1} (\bibinfo{date}{July} \bibinfo{year}{2023}), \bibinfo{pages}{4335}.
\newblock
\showISSN{2041-1723}
\urldef\tempurl%
\url{https://doi.org/10.1038/s41467-023-39850-2}
\showDOI{\tempurl}
\newblock
\shownote{Publisher: Nature Publishing Group}.


\bibitem[Varga et~al\mbox{.}(2018)]%
        {varga_designing_2018}
\bibfield{author}{\bibinfo{person}{Virag Varga}, \bibinfo{person}{Marc Wyss}, \bibinfo{person}{Gergely Vakulya}, \bibinfo{person}{Alanson Sample}, {and} \bibinfo{person}{Thomas~R. Gross}.} \bibinfo{year}{2018}\natexlab{}.
\newblock \showarticletitle{Designing {Groundless} {Body} {Channel} {Communication} {Systems}: {Performance} and {Implications}}. In \bibinfo{booktitle}{\emph{Proceedings of the 31st {Annual} {ACM} {Symposium} on {User} {Interface} {Software} and {Technology}}} \emph{(\bibinfo{series}{{UIST} '18})}. \bibinfo{publisher}{Association for Computing Machinery}, \bibinfo{address}{New York, NY, USA}, \bibinfo{pages}{683--695}.
\newblock
\showISBNx{978-1-4503-5948-1}
\urldef\tempurl%
\url{https://doi.org/10.1145/3242587.3242622}
\showDOI{\tempurl}


\bibitem[Vasisht et~al\mbox{.}(2018)]%
        {vasisht_-body_2018}
\bibfield{author}{\bibinfo{person}{Deepak Vasisht}, \bibinfo{person}{Guo Zhang}, \bibinfo{person}{Omid Abari}, \bibinfo{person}{Hsiao-Ming Lu}, \bibinfo{person}{Jacob Flanz}, {and} \bibinfo{person}{Dina Katabi}.} \bibinfo{year}{2018}\natexlab{}.
\newblock \showarticletitle{In-body backscatter communication and localization}. In \bibinfo{booktitle}{\emph{Proceedings of the 2018 {Conference} of the {ACM} {Special} {Interest} {Group} on {Data} {Communication}}} \emph{(\bibinfo{series}{{SIGCOMM} '18})}. \bibinfo{publisher}{Association for Computing Machinery}, \bibinfo{address}{New York, NY, USA}, \bibinfo{pages}{132--146}.
\newblock
\showISBNx{978-1-4503-5567-4}
\urldef\tempurl%
\url{https://doi.org/10.1145/3230543.3230565}
\showDOI{\tempurl}


\bibitem[Wang et~al\mbox{.}(2023b)]%
        {wang_locating_2023}
\bibfield{author}{\bibinfo{person}{Jingxian Wang}, \bibinfo{person}{Junbo Zhang}, \bibinfo{person}{Ke Li}, \bibinfo{person}{Chengfeng Pan}, \bibinfo{person}{Carmel Majidi}, {and} \bibinfo{person}{Swarun Kumar}.} \bibinfo{year}{2023}\natexlab{b}.
\newblock \showarticletitle{Locating {Everyday} {Objects} {Using} {NFC} {Textiles}}.
\newblock \bibinfo{journal}{\emph{Commun. ACM}} \bibinfo{volume}{66}, \bibinfo{number}{10} (\bibinfo{year}{2023}), \bibinfo{pages}{107--114}.
\newblock
\showISSN{0001-0782}
\urldef\tempurl%
\url{https://doi.org/10.1145/3615450}
\showDOI{\tempurl}


\bibitem[Wang et~al\mbox{.}(2023a)]%
        {wang_rf-sifter_2023}
\bibfield{author}{\bibinfo{person}{Xiong Wang}, \bibinfo{person}{Jun Huang}, \bibinfo{person}{Bizhao Shi}, \bibinfo{person}{Zhe Ou}, \bibinfo{person}{Guojie Luo}, \bibinfo{person}{Linghe Kong}, \bibinfo{person}{Daqing Zhang}, {and} \bibinfo{person}{Chenren Xu}.} \bibinfo{year}{2023}\natexlab{a}.
\newblock \showarticletitle{{RF}-{SIFTER}: {Sifting} {Signals} at {Layer}-0.5 to {Mitigate} {Wideband} {Cross}-{Technology} {Interference} for {IoT}}. In \bibinfo{booktitle}{\emph{Proceedings of the 29th {Annual} {International} {Conference} on {Mobile} {Computing} and {Networking}}} \emph{(\bibinfo{series}{{ACM} {MobiCom} '23})}. \bibinfo{publisher}{Association for Computing Machinery}, \bibinfo{address}{New York, NY, USA}, \bibinfo{pages}{1--14}.
\newblock
\showISBNx{978-1-4503-9990-6}
\urldef\tempurl%
\url{https://doi.org/10.1145/3570361.3592513}
\showDOI{\tempurl}


\bibitem[Whitmire et~al\mbox{.}(2019)]%
        {whitmire_aura_2019}
\bibfield{author}{\bibinfo{person}{Eric Whitmire}, \bibinfo{person}{Farshid Salemi~Parizi}, {and} \bibinfo{person}{Shwetak Patel}.} \bibinfo{year}{2019}\natexlab{}.
\newblock \showarticletitle{Aura: {Inside}-out {Electromagnetic} {Controller} {Tracking}}. In \bibinfo{booktitle}{\emph{Proceedings of the 17th {Annual} {International} {Conference} on {Mobile} {Systems}, {Applications}, and {Services}}} \emph{(\bibinfo{series}{{MobiSys} '19})}. \bibinfo{publisher}{Association for Computing Machinery}, \bibinfo{address}{New York, NY, USA}, \bibinfo{pages}{300--312}.
\newblock
\showISBNx{978-1-4503-6661-8}
\urldef\tempurl%
\url{https://doi.org/10.1145/3307334.3326090}
\showDOI{\tempurl}


\bibitem[Wicaksono et~al\mbox{.}(2020)]%
        {wicaksono_tailored_2020}
\bibfield{author}{\bibinfo{person}{Irmandy Wicaksono}, \bibinfo{person}{Carson~I. Tucker}, \bibinfo{person}{Tao Sun}, \bibinfo{person}{Cesar~A. Guerrero}, \bibinfo{person}{Clare Liu}, \bibinfo{person}{Wesley~M. Woo}, \bibinfo{person}{Eric~J. Pence}, {and} \bibinfo{person}{Canan Dagdeviren}.} \bibinfo{year}{2020}\natexlab{}.
\newblock \showarticletitle{A tailored, electronic textile conformable suit for large-scale spatiotemporal physiological sensing in vivo}.
\newblock \bibinfo{journal}{\emph{npj Flexible Electronics}} \bibinfo{volume}{4}, \bibinfo{number}{1} (\bibinfo{date}{April} \bibinfo{year}{2020}), \bibinfo{pages}{1--13}.
\newblock
\showISSN{2397-4621}
\urldef\tempurl%
\url{https://doi.org/10.1038/s41528-020-0068-y}
\showDOI{\tempurl}
\newblock
\shownote{Publisher: Nature Publishing Group}.


\bibitem[Wu et~al\mbox{.}(2025)]%
        {wu_heart_2025}
\bibfield{author}{\bibinfo{person}{Jiaying Wu}, \bibinfo{person}{Chuyu Wang}, \bibinfo{person}{Dongxu Huang}, \bibinfo{person}{Jingyi Ning}, {and} \bibinfo{person}{Lei Xie}.} \bibinfo{year}{2025}\natexlab{}.
\newblock \showarticletitle{Heart {Rate} {Variability} {Estimation} {Based} on {RFID} {Tag}-{Pair} in {Dynamic} {Environments}}.
\newblock \bibinfo{journal}{\emph{ACM Trans. Comput. Healthcare}} \bibinfo{volume}{6}, \bibinfo{number}{1} (\bibinfo{year}{2025}), \bibinfo{pages}{11:1--11:26}.
\newblock
\urldef\tempurl%
\url{https://doi.org/10.1145/3691355}
\showDOI{\tempurl}


\bibitem[Xie et~al\mbox{.}(2024)]%
        {xie_enabling_2024}
\bibfield{author}{\bibinfo{person}{Mingqi Xie}, \bibinfo{person}{Meng Jin}, \bibinfo{person}{Fengyuan Zhu}, \bibinfo{person}{Yuzhe Zhang}, \bibinfo{person}{Xiaohua Tian}, \bibinfo{person}{Xinbing Wang}, {and} \bibinfo{person}{Chenghu Zhou}.} \bibinfo{year}{2024}\natexlab{}.
\newblock \showarticletitle{Enabling {High}-rate {Backscatter} {Sensing} at {Scale}}. In \bibinfo{booktitle}{\emph{Proceedings of the 30th {Annual} {International} {Conference} on {Mobile} {Computing} and {Networking}}} \emph{(\bibinfo{series}{{ACM} {MobiCom} '24})}. \bibinfo{publisher}{Association for Computing Machinery}, \bibinfo{address}{New York, NY, USA}, \bibinfo{pages}{124--138}.
\newblock
\showISBNx{9798400704895}
\urldef\tempurl%
\url{https://doi.org/10.1145/3636534.3649351}
\showDOI{\tempurl}


\bibitem[Ye et~al\mbox{.}(2022)]%
        {ye_body-centric_2022}
\bibfield{author}{\bibinfo{person}{Huizhong Ye}, \bibinfo{person}{Chi-Jung Lee}, \bibinfo{person}{Te-Yen Wu}, \bibinfo{person}{Xing-Dong Yang}, \bibinfo{person}{Bing-Yu Chen}, {and} \bibinfo{person}{Rong-Hao Liang}.} \bibinfo{year}{2022}\natexlab{}.
\newblock \showarticletitle{Body-{Centric} {NFC}: {Body}-{Centric} {Interaction} with {NFC} {Devices} {Through} {Near}-{Field} {Enabled} {Clothing}}. In \bibinfo{booktitle}{\emph{Proceedings of the 2022 {ACM} {Designing} {Interactive} {Systems} {Conference}}} \emph{(\bibinfo{series}{{DIS} '22})}. \bibinfo{publisher}{Association for Computing Machinery}, \bibinfo{address}{New York, NY, USA}, \bibinfo{pages}{1626--1639}.
\newblock
\showISBNx{978-1-4503-9358-4}
\urldef\tempurl%
\url{https://doi.org/10.1145/3532106.3534569}
\showDOI{\tempurl}


\bibitem[Yu et~al\mbox{.}(2022)]%
        {yu_magnetoelectric_2022}
\bibfield{author}{\bibinfo{person}{Zhanghao Yu}, \bibinfo{person}{Fatima~T. Alrashdan}, \bibinfo{person}{Wei Wang}, \bibinfo{person}{Matthew Parker}, \bibinfo{person}{Xinyu Chen}, \bibinfo{person}{Frank~Y. Chen}, \bibinfo{person}{Joshua Woods}, \bibinfo{person}{Zhiyu Chen}, \bibinfo{person}{Jacob~T. Robinson}, {and} \bibinfo{person}{Kaiyuan Yang}.} \bibinfo{year}{2022}\natexlab{}.
\newblock \showarticletitle{Magnetoelectric backscatter communication for millimeter-sized wireless biomedical implants}. In \bibinfo{booktitle}{\emph{Proceedings of the 28th {Annual} {International} {Conference} on {Mobile} {Computing} {And} {Networking}}} \emph{(\bibinfo{series}{{MobiCom} '22})}. \bibinfo{publisher}{Association for Computing Machinery}, \bibinfo{address}{New York, NY, USA}, \bibinfo{pages}{432--445}.
\newblock
\showISBNx{978-1-4503-9181-8}
\urldef\tempurl%
\url{https://doi.org/10.1145/3495243.3560541}
\showDOI{\tempurl}


\bibitem[Zargham and Gulak(2012)]%
        {zargham_maximum_2012}
\bibfield{author}{\bibinfo{person}{Meysam Zargham} {and} \bibinfo{person}{P.~Glenn Gulak}.} \bibinfo{year}{2012}\natexlab{}.
\newblock \showarticletitle{Maximum {Achievable} {Efficiency} in {Near}-{Field} {Coupled} {Power}-{Transfer} {Systems}}.
\newblock \bibinfo{journal}{\emph{IEEE Transactions on Biomedical Circuits and Systems}} \bibinfo{volume}{6}, \bibinfo{number}{3} (\bibinfo{date}{June} \bibinfo{year}{2012}), \bibinfo{pages}{228--245}.
\newblock
\showISSN{1940-9990}
\urldef\tempurl%
\url{https://doi.org/10.1109/TBCAS.2011.2174794}
\showDOI{\tempurl}
\newblock
\shownote{Conference Name: IEEE Transactions on Biomedical Circuits and Systems}.


\bibitem[Zhang et~al\mbox{.}(2023)]%
        {zhang_nfcapsule_2023}
\bibfield{author}{\bibinfo{person}{Junbo Zhang}, \bibinfo{person}{Gaurav Balakrishnan}, \bibinfo{person}{Sruti Srinidhi}, \bibinfo{person}{Arnav Bhat}, \bibinfo{person}{Swarun Kumar}, {and} \bibinfo{person}{Christopher Bettinger}.} \bibinfo{year}{2023}\natexlab{}.
\newblock \showarticletitle{{NFCapsule}: {An} {Ingestible} {Sensor} {Pill} for {Eosinophilic} {Esophagitis} {Detection} {Based} on near-{Field} {Coupling}}. In \bibinfo{booktitle}{\emph{Proceedings of the 20th {ACM} {Conference} on {Embedded} {Networked} {Sensor} {Systems}}} \emph{(\bibinfo{series}{{SenSys} '22})}. \bibinfo{publisher}{Association for Computing Machinery}, \bibinfo{address}{New York, NY, USA}, \bibinfo{pages}{75--90}.
\newblock
\showISBNx{978-1-4503-9886-2}
\urldef\tempurl%
\url{https://doi.org/10.1145/3560905.3568523}
\showDOI{\tempurl}


\bibitem[Zhang et~al\mbox{.}(2016)]%
        {zhang_hitchhike_2016}
\bibfield{author}{\bibinfo{person}{Pengyu Zhang}, \bibinfo{person}{Dinesh Bharadia}, \bibinfo{person}{Kiran Joshi}, {and} \bibinfo{person}{Sachin Katti}.} \bibinfo{year}{2016}\natexlab{}.
\newblock \showarticletitle{{HitchHike}: {Practical} {Backscatter} {Using} {Commodity} {WiFi}}. In \bibinfo{booktitle}{\emph{Proceedings of the 14th {ACM} {Conference} on {Embedded} {Network} {Sensor} {Systems} {CD}-{ROM}}} \emph{(\bibinfo{series}{{SenSys} '16})}. \bibinfo{publisher}{Association for Computing Machinery}, \bibinfo{address}{New York, NY, USA}, \bibinfo{pages}{259--271}.
\newblock
\showISBNx{978-1-4503-4263-6}
\urldef\tempurl%
\url{https://doi.org/10.1145/2994551.2994565}
\showDOI{\tempurl}


\bibitem[ZHANG et~al\mbox{.}(2016)]%
        {zhang_enabling_2016}
\bibfield{author}{\bibinfo{person}{PENGYU ZHANG}, \bibinfo{person}{Mohammad Rostami}, \bibinfo{person}{Pan Hu}, {and} \bibinfo{person}{Deepak Ganesan}.} \bibinfo{year}{2016}\natexlab{}.
\newblock \showarticletitle{Enabling {Practical} {Backscatter} {Communication} for {On}-body {Sensors}}. In \bibinfo{booktitle}{\emph{Proceedings of the 2016 {ACM} {SIGCOMM} {Conference}}} \emph{(\bibinfo{series}{{SIGCOMM} '16})}. \bibinfo{publisher}{Association for Computing Machinery}, \bibinfo{address}{New York, NY, USA}, \bibinfo{pages}{370--383}.
\newblock
\showISBNx{978-1-4503-4193-6}
\urldef\tempurl%
\url{https://doi.org/10.1145/2934872.2934901}
\showDOI{\tempurl}


\bibitem[Zhao et~al\mbox{.}(2020)]%
        {zhao_nfc_2020}
\bibfield{author}{\bibinfo{person}{Renjie Zhao}, \bibinfo{person}{Purui Wang}, \bibinfo{person}{Yunfei Ma}, \bibinfo{person}{Pengyu Zhang}, \bibinfo{person}{Hongqiang~Harry Liu}, \bibinfo{person}{Xianshang Lin}, \bibinfo{person}{Xinyu Zhang}, \bibinfo{person}{Chenren Xu}, {and} \bibinfo{person}{Ming Zhang}.} \bibinfo{year}{2020}\natexlab{}.
\newblock \showarticletitle{{NFC}+: {Breaking} {NFC} {Networking} {Limits} through {Resonance} {Engineering}}. In \bibinfo{booktitle}{\emph{Proceedings of the {Annual} conference of the {ACM} {Special} {Interest} {Group} on {Data} {Communication} on the applications, technologies, architectures, and protocols for computer communication}} \emph{(\bibinfo{series}{{SIGCOMM} '20})}. \bibinfo{publisher}{Association for Computing Machinery}, \bibinfo{address}{New York, NY, USA}, \bibinfo{pages}{694--707}.
\newblock
\showISBNx{978-1-4503-7955-7}
\urldef\tempurl%
\url{https://doi.org/10.1145/3387514.3406219}
\showDOI{\tempurl}


\bibitem[Zhao et~al\mbox{.}(2015)]%
        {zhao_nfc-wisp_2015}
\bibfield{author}{\bibinfo{person}{Yi Zhao}, \bibinfo{person}{Joshua~R. Smith}, {and} \bibinfo{person}{Alanson Sample}.} \bibinfo{year}{2015}\natexlab{}.
\newblock \showarticletitle{{NFC}-{WISP}: {A} sensing and computationally enhanced near-field {RFID} platform}. In \bibinfo{booktitle}{\emph{2015 {IEEE} {International} {Conference} on {RFID} ({RFID})}}. \bibinfo{pages}{174--181}.
\newblock
\urldef\tempurl%
\url{https://doi.org/10.1109/RFID.2015.7113089}
\showDOI{\tempurl}
\newblock
\shownote{ISSN: 2374-0221}.


\bibitem[Zheng et~al\mbox{.}(2021)]%
        {zheng_more-fi_2021}
\bibfield{author}{\bibinfo{person}{Tianyue Zheng}, \bibinfo{person}{Zhe Chen}, \bibinfo{person}{Shujie Zhang}, \bibinfo{person}{Chao Cai}, {and} \bibinfo{person}{Jun Luo}.} \bibinfo{year}{2021}\natexlab{}.
\newblock \showarticletitle{{MoRe}-{Fi}: {Motion}-robust and {Fine}-grained {Respiration} {Monitoring} via {Deep}-{Learning} {UWB} {Radar}}. In \bibinfo{booktitle}{\emph{Proceedings of the 19th {ACM} {Conference} on {Embedded} {Networked} {Sensor} {Systems}}} \emph{(\bibinfo{series}{{SenSys} '21})}. \bibinfo{publisher}{Association for Computing Machinery}, \bibinfo{address}{New York, NY, USA}, \bibinfo{pages}{111--124}.
\newblock
\showISBNx{978-1-4503-9097-2}
\urldef\tempurl%
\url{https://doi.org/10.1145/3485730.3485932}
\showDOI{\tempurl}


\bibitem[Zhu et~al\mbox{.}(2024)]%
        {zhu_robust_2024}
\bibfield{author}{\bibinfo{person}{Xia Zhu}, \bibinfo{person}{Ke Wu}, \bibinfo{person}{Xiaohang Xie}, \bibinfo{person}{Stephan~W. Anderson}, {and} \bibinfo{person}{Xin Zhang}.} \bibinfo{year}{2024}\natexlab{}.
\newblock \showarticletitle{A robust near-field body area network based on coaxially-shielded textile metamaterial}.
\newblock \bibinfo{journal}{\emph{Nature Communications}} \bibinfo{volume}{15}, \bibinfo{number}{1} (\bibinfo{date}{Aug.} \bibinfo{year}{2024}), \bibinfo{pages}{6589}.
\newblock
\showISSN{2041-1723}
\urldef\tempurl%
\url{https://doi.org/10.1038/s41467-024-51061-x}
\showDOI{\tempurl}
\newblock
\shownote{Publisher: Nature Publishing Group}.


\end{thebibliography}

\appendix

\end{document}